\documentclass{amsart}
\usepackage{geometry}
\usepackage{graphics}
\usepackage[pdftex]{hyperref}
\usepackage[dvips]{graphicx}
\usepackage{pgf}
\usepackage{tikz}
\usetikzlibrary{arrows,automata,positioning}

\tikzset{main node/.style={circle,fill=blue!20,draw,minimum size=1cm,inner sep=0pt},
            }

\newtheorem{theorem}{Theorem}[section]
\newtheorem{lemma}[theorem]{Lemma}
\newtheorem{pro}[theorem]{Proposition}
\newtheorem{cor}[theorem]{Corollary}
\newtheorem{remark}[theorem]{Remark}

\theoremstyle{definition}
\newtheorem{definition}[theorem]{Definition}
\newtheorem{example}[theorem]{Example}

\theoremstyle{remark}

\numberwithin{equation}{section}

\begin{document}
\pagestyle{plain}

\author{Carlos F. Lardizabal}

\address{Instituto de Matem\'atica e Estat\'istica - Universidade Federal do Rio Grande do Sul - UFRGS - Av. Bento Gon\c calves 9500 - CEP 91509-900 Porto Alegre, RS, Brazil}
\email{cfelipe@mat.ufrgs.br}

\def\laa{\langle}
\def\raa{\rangle}
\def\qed{\begin{flushright} $\square$ \end{flushright}}
\def\qee{\begin{flushright} $\Diamond$ \end{flushright}}
\def\ov{\overline}

\title{Open Quantum Random Walks and the Mean Hitting Time Formula}

\begin{abstract} We make use of the Open Quantum Random Walk setting due to S. Attal, F. Petruccione, C. Sabot and I. Sinayskiy [J. Stat. Phys. (2012) 147:832-852] in order to discuss hitting times and a quantum version of the Mean Hitting Time Formula from  classical probability theory. We study an open quantum notion of hitting probability on a finite collection of sites and with this we are able to describe the problem in terms of linear maps and its matrix representations. After setting an open quantum version of the fundamental matrix for ergodic Markov chains we are able to prove our main result and as consequence a  version of the Random Target Lemma. We also study a mean hitting time formula in terms of the minimal polynomial associated to the matrix representation of the quantum walk. We discuss applications of the results to open quantum dynamics on graphs together with open questions.
\end{abstract}

\date{\today}

\maketitle


\section{Introduction}

In this work we are interested in certain probabilistic aspects of open quantum systems \cite{breuerp}. Informally, the term open means that the system of interest is not isolated from the environment and as such it is subject to interference. Then, instead of having a unitary evolution (as in ideal quantum dynamics), we are presented with a dissipative system which is described by certain Lindblad equations \cite{breuerp,alicki,nielsen} and its solutions given by completely positive maps (quantum channels) \cite{bhatia,petz}.

\medskip

In the search of quantum counterparts of classical probabilistic notions, one often notices that there may be more than one reasonable definition. For instance, the notion of recurrence for quantum systems has received considerable attention in mathematical physics and in quantum probability literature as several constructions exist, see e.g., \cite{accardi,fagnola,fagnola2,werner,ls2014,coliveira,stefanak}, each with their own characteristics depending, for instance, on the Hilbert space of interest. In general these are nonequivalent notions, meaning that a system may be recurrent in one sense but not in another, see e.g. the discussion in \cite{werner}.

\medskip

In the present work we examine a notion of hitting times and hitting probabilities for certain classes of open quantum systems. We will have in mind the dynamics of a quantum particle on a graph and then ask for the time of first visit of such particle to a given vertex, subject to some kind of monitoring. Besides describing a mathematical setting for such study, it is one of our goals to explain what can be learned from a quantum system if we perform the computation of hitting times and its related averages. Our starting point will be a known construction in classical probability.

\medskip

Let us consider the classical problem of calculating the average time of first visit to a certain site. Consider a finite state irreducible Markov chain $(X_n:n=0,1,\dots)$ associated to a stochastic matrix $P=(p_{ij})$. Let
\begin{equation}\label{counter0}
T_i:=\min\{n\geq 0: X_n=i\}
\end{equation}
and
\begin{equation}
Z_{ij}:=\sum_{n=0}^\infty (p_{ij}^{(n)}-\pi_j),
\end{equation}
where $\pi=(\pi_i)$ is the unique stationary distribution for $P$. We call $Z=(Z_{ij})$ the fundamental matrix. Then the Mean Hitting Time Formula relates these elements in the following way:
\begin{equation}\label{mhtf_classica}
\pi_jE_iT_j=Z_{jj}-Z_{ij}
\end{equation}
This result has a number of consequences, one of them being the Random Target Lemma: for all $i$,
\begin{equation}\label{rtl_classica}
\sum_j \pi_j E_iT_j=\sum_j Z_{jj},
\end{equation}
so this quantity does not depend on $i$. Probabilistic proofs of these results can be seen in \cite{aldous}. In \cite{bremaud}, a different proof of (\ref{mhtf_classica}) is presented, relying on probability and matrix analysis arguments.

\medskip

Then, we may ask: can we obtain a quantum version of the Mean Hitting Time Formula? If so, what kind of quantum information is obtained from it? First and foremost, one should consider a setting where a reasonable, quantum notion of hitting time is available. In this work, we will discuss such questions in the context of Open Quantum Random Walks (OQWs). OQWs, a quantum generalization of Markov chains, have been first discussed by S. Attal et al. \cite{attal}. We recall the basic construction. Let $\{B_{ij}\}_{i,j=1,\dots,k}$ belong to $M_n(\mathbb{C})$, the order $n$ complex matrices, such that for each $j=1,\dots,k$,
\begin{equation}
\sum_{i=1}^k B_{ij}^*B_{ij}=I,
\end{equation}
where $I$ denotes the order $n$ identity matrix. We say that $k$ is the number of {\bf sites} and $n$ is the {\bf degree of freedom} on each site. Define
\begin{equation}\label{dens_attal}
\rho:=\sum_{i=1}^k\rho_i\otimes |i\rangle\langle i|,\;\;\;\rho_i\in M_n(\mathbb{C}), \;\;\;\rho_i\geq 0, \;\;\;\sum_{i=1}^k Tr(\rho_i)=1,
\end{equation}
where $\rho_i\geq 0$ means that $\rho_i$ is positive semidefinite. For a given initial density matrix  of such form, the {\bf Open Quantum Random Walk} (OQW) on $k$ sites induced by the $B_{ij}$, $i,j=1,\dots,k$ is, by definition \cite{attal}, the map
\begin{equation}\label{oqrw_def}
\Phi(\rho):=\sum_{i=1}^k\Big(\sum_{j=1}^k B_{ij}\rho_j B_{ij}^{*}\Big)\otimes |i\rangle\langle i|.
\end{equation}
Now we remark that density matrices of the form (\ref{dens_attal}) are preserved under the action of OQWs. That is, given one such density we have that $\Phi(\rho)$ is also a summation of terms of the form $\eta_i\otimes|i\rangle\langle i|$, $\eta_i\geq 0$, $\sum_i Tr(\eta_i)=1$, see  \cite{attal}. In particular, the sites $i=|i\rangle$ serve as an index for the entries in a given vector. We say that $B_{ij}$ is the effect matrix of transition from site $j$ to site $i$. If the state of the chain at time $n$ is $\rho^{(n)}=\rho\otimes|j\rangle\langle j|$ then at time $n+1$ it jumps to
\begin{equation}
\rho^{(n+1)}=\frac{B_{ij}\rho B_{ij}^*}{p(j,i)}\otimes |i\rangle\langle i|,\;\;\;i\in\mathbb{Z}
\end{equation}
with probability
\begin{equation}\label{basprobfor}
p(j,i)=Tr(B_{ij}\rho B_{ij}^{*})
\end{equation}
This is a well-known transition rule seen in quantum mechanics, which depends on a density matrix. In probability notation, we have a homogeneous Markov chain $(\rho_n,X_n)$ with values in $D_n\times \{1,\dots,k\}$ (continuous state, discrete time), where $D_n$ consists of the set of density matrices on $M_n(\mathbb{C})$ (see \cite{attal,ls2015}), satisfying: from any position $(\rho,j)$ one jumps to
\begin{equation}\label{qtapp}
\Bigg(\frac{B_{ij}\rho B_{ij}^*}{p(j,i)},i\Bigg)
\end{equation}
with probability given by (\ref{basprobfor}). This is the {\bf quantum trajectories} formalism of OQWs. Informally, we have an open quantum process for which we perform measurements at each time step, that is, we have a monitored procedure. Also it is worth noting that the dynamics of OQWs are quite different from the usual (closed) quantum random walks, we refer the reader to \cite{portugal} and \cite{salvador} for more on this kind of walk.

\medskip

Since the publication of \cite{attal}, several articles have appeared concerning probabilistic and statistical properties of OQWs. For instance, \cite{attal2} discusses a Central Limit Theorem for OQWs; in \cite{carbone}, Carbone and Pautrat discuss reducibility, periodicity and ergodicity properties and in \cite{carbone2} a large deviation principle is presented. In \cite{konno}, Konno and Yoo present limit theorems in terms of an integral formula obtained by the Fourier transform of the evolving density matrix. Continuous time OQWs are studied by Pellegrini \cite{pellegrini}. In \cite{sinayskiy2}, Petruccione and Sinayskiy discuss the microscopic derivation of OQWs. It is shown that we can obtain Kraus operators from the continuous time generator via a discretization procedure so one can consider an iterative evolution. In \cite{sinayskiy3} it is discussed the OQW implementation of dissipative quantum computing algorithms (the latter two topics are further discussed later in this work in relation with hitting times).

\medskip

 The problem of site recurrence of OQWs is studied in \cite{ls2014} and hitting times for OQWs are first discussed in \cite{ls2015}, where open quantum versions of the gambler's ruin and birth-and-death chains are examined, and where the ergodicity of sequences of OQWs is considered. Also see \cite{liu,pawela,sinayskiy,sinayskiy4,xiong} for further applications.

\section{Hitting Times for OQWs and Statement of Results}

Hitting times are a central object in the theory of Markov chains \cite{aldous,bremaud,chenz,grimmett,rimfo2}, so we may be interested in the differences (and similarities) between classical and quantum settings. Motivated by a notion of recurrence of OQWs presented in \cite{ls2014}, a notion of hitting time for OQWs is proposed in \cite{ls2015}. We recall these constructions in finite dimension in what follows. Then we work towards the necessary objects needed to prove an OQW version of the Mean Hitting Time Formula. After this notion is established, we are able to obtain a basic formalism of quantum hitting times and have it developed in a manner which may aid the study of problems in quantum information theory. This is made clear in the examples, and throughout this work, see Section \ref{openq}. Now we address the following points:

\bigskip

1. We need a notion of hitting probability for OQWs. We adopt the following, described in \cite{ls2015}.

\begin{definition} The {\bf probability of first visit to site $i$ at time $r$}, starting at $\rho_j\otimes|j\rangle\langle j|$ is denoted by $b_r(\rho_j;i)$. This is the sum of the traces of all paths starting at $\rho_j\otimes |j\rangle\langle j|$ and reaching $i$ for the first time at the $r$-th step. The probability starting from $\rho_j\otimes |j\rangle\langle j|$ that the walk ever hits site $i$ is
\begin{equation}
h_{ij}(\rho_j):=\sum_{r=0}^\infty b_r(\rho_j;i),\;\;\;i\neq j
\end{equation}
and $h_{jj}(\rho_j):=1$. This is the {\bf probability of visiting site $i$}, given that the walk started at site $j$.
\end{definition}

\begin{definition} For fixed initial state and final site, the {\bf mean hitting time} is
\begin{equation}
k_{ij}(\rho_j):=\sum_{r=1}^\infty r\;b_r(\rho_j;i).
\end{equation}
\end{definition}

\medskip

2. A second step is to write expressions of hitting times in matrix terms. Let $\pi(i\leftarrow j;r)$ be the set of all products of matrices corresponding to the sequences of sites that a walk is allowed to perform with $\Phi$, beginning at site $|j\rangle$, first reaching site $|i\rangle$ in $r$ steps. For instance, reading matrices and indices from right to left, we have $B_{43}B_{32}B_{21}B_{12}B_{21}\in\pi(4\leftarrow 1;5)$ as this corresponds to moving right, moving left and then moving right 3 times. Let $\pi(i\leftarrow j)=\cup_r \pi(i\leftarrow j;r)$. Then, by definition,
\begin{equation}
b_r(\rho_j;i)=\sum_{C\in \pi(i\leftarrow j;r)} Tr(C\rho_j C^*)
\end{equation}
and
\begin{equation}\label{defhitt}
h_{ij}(\rho_j)=\sum_{r=0}^\infty b_r(\rho_j;i)=\sum_{r=0}^\infty\sum_{C\in \pi(i\leftarrow j;r)} Tr(C\rho_j C^*)=\sum_{C\in \pi(i\leftarrow j)} Tr(C\rho_j C^*)
\end{equation}
If $C\in M_n(\mathbb{C})$, define $M_C:M_n(\mathbb{C})\to M_n(\mathbb{C})$,
\begin{equation}\label{defconjug}
M_C(X):=CXC^*,\;\;\;X\in M_n(\mathbb{C})
\end{equation}
For simplicity, assume $k=2$. Let
\begin{equation}
\hat{H}=\begin{bmatrix} \hat{h}_{11} & \hat{h}_{12} \\ \hat{h}_{21} & \hat{h}_{22}\end{bmatrix}:=\begin{bmatrix} \sum_{C\in\pi(1\leftarrow 1)}M_C & \sum_{C\in\pi(1\leftarrow 2)}M_C\\
\sum_{C\in\pi(2\leftarrow 1)}M_C & \sum_{C\in\pi(2\leftarrow 2)}M_C\end{bmatrix}
\end{equation}
We give the analogous definition for $k>2$ sites. In words, $\hat{h}_{ij}$ denotes the map associated with summing all possible ways of going from $|j\rangle$ to $|i\rangle$. Then for every density $\rho_j$ we have, by the definition of $\hat{h}_{ij}$,
\begin{equation}
h_{ij}(\rho_j)=Tr(\hat{h}_{ij}\rho_j)
\end{equation}
In a similar way as in the definition for the hitting probability operator $\hat{H}$, define the mean hitting time operator $\hat{K}=(\hat{k}_{ij})$,
\begin{equation}\label{k_eq1}
\hat{K}=\begin{bmatrix}\hat{k}_{11} & \hat{k}_{12} \\ \hat{k}_{21} & \hat{k}_{22}\end{bmatrix}:=\begin{bmatrix} \sum_r r\sum_{C\in\pi(1\leftarrow 1;r)}M_C & \sum_r r\sum_{C\in\pi(1\leftarrow 2;r)}M_C \\ \sum_r r\sum_{C\in\pi(2\leftarrow 1;r)}M_C & \sum_r r\sum_{C\in\pi(2\leftarrow 2;r)}M_C\end{bmatrix}
\end{equation}
and note that
\begin{equation}
k_{ij}(\rho_j)=Tr(\hat{k}_{ij}\rho_j),\;\;\; i\neq j,
\end{equation}
and we set $k_{ii}(\rho)=0$ for every $\rho$ density, since we begin counting visits at time zero, as in eq. (\ref{counter0}). We give the analogous definition for $k>2$ sites.

\medskip

3. It is then necessary to define an OQW version of the fundamental matrix of ergodic Markov chains \cite{aldous,bremaud}. Let $\hat{\Omega}$ denote the block matrix where each block entry equals $\hat{\Omega}_{kl}=\frac{1}{kn}\sum_{i,j=1}^nE_{ij}\otimes E_{ij}$, for $k,l=1,\dots,n$ and $E_{ij}\in M_n(\mathbb{C})$ are the matrix units: $(E_{ij})_{rs}=1$ if $(r,s)=(i,j)$, and equals zero otherwise. Note that $\hat{\Omega}^2=\hat{\Omega}$. The idea behind this definition is that there are OQWs such that the iterates of its block representation converge to $\hat{\Omega}$. This should be compared with the fact that the columns (rows) of a stochastic matrix corresponding to finite irreducible aperiodic Markov chains converge to the unique stationary probability vector. Properties on $\hat{\Omega}$ will be discussed later in this work. Let
\begin{equation}
\mathcal{E}:=\{\Phi \; OQW: \hat{\Phi}^r\to \hat{\Omega},\; r\to\infty\},
\end{equation}
where the convergence is in the sense of entrywise convergence of the associated matrix representations (to be discussed later). We call $\mathcal{E}$ the collection of {\bf ergodic} OQWs, which is the set of OQWs such that its iterates converge to a unique asymptotic limit, described by $\hat{\Omega}$, regardless of its initial state. Then we can define:

\medskip

\begin{definition}
(Fundamental matrix of ergodic OQWs). For every $\Phi\in\mathcal{E}$ acting on a finite collection of sites, define
\begin{equation}\label{fund_mat}
\hat{\mathcal{Z}}:=\hat{I}+\sum_{r\geq 1}(\hat{\Phi}^r-\hat{\Omega})=(\hat{I}-\hat{\Phi}+\hat{\Omega})^{-1}
\end{equation}
\end{definition}
That this definition makes sense in the set $\mathcal{E}$ will be discussed in the following sections. See also \cite{szehr} for a related discussion on this map in the context of quantum channels. We may state the main result of this work.

\begin{theorem}\label{quantum_aldous3}(Mean Hitting Time Formula for OQWs). Let $\Phi$ denote a finite ergodic OQW with degree of freedom $n\geq 2$ and $k\geq 2$ sites and let $\hat{\mathcal{Z}}$ denote its fundamental matrix. Let $\hat{D}$ be the diagonal matrix operator with diagonal entries $\hat{k}_{ii}$ and let $\hat{N}:=\hat{K}-\hat{D}$. Then for every $\rho$ density matrix, for all $i,j=1,\dots, k$,
\begin{equation}\label{aldous_eqq1}
Tr(\hat{N}_{ij}\rho)=Tr([(\hat{D}\hat{\mathcal{Z}})_{ii}-(\hat{D}\hat{\mathcal{Z}})_{ij}]\rho)
\end{equation}
\end{theorem}

\begin{cor}
If $\hat{D}$ is invertible and there is $c$ such that $Tr(\hat{k}_{ii}\eta)=cTr(\eta)$, all $i, \eta$, then
\begin{equation}\label{aldous_eqnova}
Tr((\hat{D}^{-1}\hat{N})_{ij}\rho)=Tr([\hat{\mathcal{Z}}_{ii}-\hat{\mathcal{Z}}_{ij}]\rho)
\end{equation}
\end{cor}
Expression (\ref{aldous_eqnova}) should be compared with (\ref{mhtf_classica}).

\begin{cor}\label{rtlemma}
(Random Target Lemma for OQWs). With the assumptions of Corollary 1, for all $j$, $\rho$ density matrix, the expression
\begin{equation}\label{rtlem11}
t_{\odot}(\rho):=\sum_i Tr((\hat{D}^{-1}\hat{N})_{ij}\rho)
\end{equation}
does not depend on $j$.
\end{cor}
Expression (\ref{rtlem11}) should be compared with (\ref{rtl_classica}). The number $t_{\odot}(\rho)$ will be called the {\bf target time} of the OQW with respect to the initial density $\rho$. The  classical and quantum interpretation of a target time is discussed in Section 3.

\medskip

Based on a result due to H. Chen and F. Zhang \cite{chenz}, we discuss another mean hitting time formula in Section \ref{sec6}. Such result makes use of the minimal polynomial of the associated stochastic matrix. In our setting the proof is a simple adaptation of the mentioned work, taking in consideration the Mean Hitting Time Formula for OQWs.

\begin{theorem}\label{chzhthm}
(Mean hitting time formula in terms of the minimal polynomial). Let $\Phi\in\mathcal{E}$ with degree of freedom $n\geq 2$ and $k\geq 2$ sites, let $\hat{D}$ be the diagonal matrix operator with diagonal entries $\hat{k}_{ii}$ and let $\hat{N}:=\hat{K}-\hat{D}$. We have, for all $\rho$ density matrix, for all $i,j=1,\dots, k$,
\begin{equation}
Tr(\hat{N}_{ij}\rho)=Tr\Big[\Big(\sum_{s=0}^\infty\Big((\hat{D}\hat{\Phi}^s)_{ii}-(\hat{D}\hat{\Phi}^s)_{ij}\Big)\rho\Big)\Big]
\end{equation}
and
\begin{equation}
\sum_{s=0}^\infty\Big((\hat{D}\hat{\Phi}^s)_{ii}-(\hat{D}\hat{\Phi}^s)_{ij}\Big)=\frac{1}{f(1)}\sum_{s=0}^{r-1}\sum_{l=0}^{r-s-1}a_l((\hat{D}\hat{\Phi}^s)_{ii}-(\hat{D}\hat{\Phi}^s)_{ij}),
\end{equation}
where $\hat{D}=diag(\hat{k}_{11},\dots,\hat{k}_{22})$ and $r$ is the degree of the polynomial $f$ satisfying $(x-1)f(x)=p(x)$, the minimal polynomial for $\Phi$.
\end{theorem}

In this work we will consider mostly OQWs with $k=2$ sites and degree of freedom given by order $n=2$ matrices as this case already presents nontrivial aspects. Generalizations of the definitions and corresponding proofs for higher dimensions are straightforward. The validity of certain statements for $n, k>2$ sites are discussed separately whenever necessary.

\section{Applications of mean hitting times to open quantum systems}\label{openq}


We note that hitting times have been extensively discussed in the setting of unitary (coined) quantum walks (see the survey \cite{salvador} for references). As it is well-known, there are notions of quantum (mean) hitting times for unitary quantum walks on finite graphs which are at least quadratically smaller than the classical hitting time of a random walk on the same graph \cite{portugal}. Then we may ask the following question in an open quantum context: what are the possible outcomes for mean hitting times for OQWs presented in this work and how does it compare with its unitary quantum counterparts?

\medskip

The reason for choosing OQWs as our model of open dynamics lies, in part, on the versatility of the model when applied to the study of dynamics on graphs, and also on its rigorous mathematical formalism.  Such formalism has at the same time a clear connection with classical probability theory (exemplified for instance by the use of martingales in the proof of the Central Limit Theorem \cite{attal2}, or by the arguments leading to a large deviation principle seen in \cite{carbone2}), but also presents a noncommutative character in terms of trace calculation of density matrices and transition rules (also see the use of well-known operator algebra results on irreducibility, seen in \cite{carbone}). Moreover, the fact that classical Markov chains are a particular case of OQWs is explained in [\cite{attal}, Section 6].

\medskip

It is worth mentioning that mean return times to states (and sites), which are important cases of mean hitting times, have been studied recently in the context of quantum channels. See, for instance,  \cite{sinkovicz} for the problem of mean return time to states for unital channels, \cite{sinkovicz2} for a generalized Kac's lemma, and \cite{ls2015} for the mean return time to sites (in the case of OQWs for which positive recurrence is valid).

\medskip

In the following we comment on other topics which, in part, justify our interest on hitting times and serve as an illustration of certain open questions and possible research directions. We also refer the reader to \cite{ls2015} for basic results and applications on hitting times for OQWs.

\medskip

{\bf 1. Walks on the $N$-path.} Consider the classical problem of walking from vertex $1$ to vertex $N$ on the $N$-path, the graph consisting of a single path of $N$ vertices. Let $k_{i,j}$, $j<i$ be the mean hitting time of reaching $i$, starting from $j$. Suppose the border sites are repelling and the remaining ones possess fair transitions ($p=1/2$). Then clearly $k_{21}=1$ and
\begin{equation}\label{class_rec}
k_{i+1,i}=\frac{1}{2}\cdot 1+\frac{1}{2}\cdot(1+k_{i,i-1}+k_{i+1,i}),\;\;\;2\leq i\leq n-1
\end{equation}
This equality reads as follows: either we move right to site $i+1$ and we are done (with probability $1/2$), or else we move left and so we need 1 unit of time (the one spent moving left) plus the mean time of moving right twice (this also occurring with probability 1/2). It is a simple matter to solve this recursion and show that the expected time for a random walk starting at $1$ to reach site $N$ is $(N-1)^2$, that is, $O(N^2)$.

\medskip

A natural question in the context of OQWs is: given $L$, $R$ matrices such that $L^*L+R^*R=I$, what is the behaviour of the mean hitting time for the open walk on the $N$-path associated to $L$ and $R$? Note that the OQW setting has the classical problem as a particular case: choose $L=R=I/\sqrt{2}$. We may rephrase our question: are there any choices of $L$ and $R$ such that they are fair (induces equal probability to both sides) and such that the mean hitting time is different from the classical case?

\begin{center}
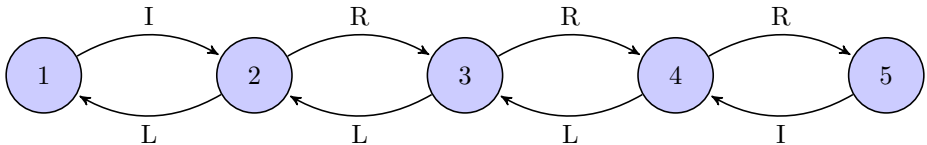
\begin{figure}[h]
\begin{tikzpicture}[->,>=stealth',shorten >=1pt,auto,node distance=2.8cm,
                    semithick]
  \tikzstyle{every state}=[circle,fill=blue!20,draw,minimum size=1cm,inner sep=0pt]

  \node[state] (B) {$1$};
  \node[state] (D) [right of=B] {$2$};
  \node[state] (E) [right of=D] {$3$};
  \node[state] (F) [right of=E] {$4$};
  \node[state] (G) [right of=F] {$5$};
  \path (B) edge   [bend left]     node {I} (D)
        (D) edge   [bend left]     node {R} (E)
        (D) edge   [bend left]     node {L} (B)
        (E) edge   [bend left]     node {R} (F)
        (E) edge   [bend left]     node {L} (D)
        (F) edge   [bend left]     node {L} (E)
        (F) edge   [bend left]     node {R} (G)
        (G) edge   [bend left]     node {I} (F)
        ;
\end{tikzpicture}
\caption{$5$-path with general transition effects $L$, $R$, $L^*L+R^*R=I$. The mean hitting time for the OQW to move from one extreme of the path to the other is an open quantum generalization of the classical problem and has a density matrix degree of freedom evolving at each iteration. Two related problems are: a) the nonhomogeneous version, for which we may have distinct transitions $L_i$, $R_i$, for each vertex $i$, and b) calculating the average return time to some given vertex.}
\end{figure}
\end{center}
Recall that
\begin{equation}
k_{ij}(\rho_j)=\sum_{r=1}^\infty r\;b_r(\rho_j;i),\;\;\;b_r(\rho_j;i)=\sum_{C\in \pi(i\leftarrow j;r)} Tr(C\rho_j C^*),
\end{equation}
so, in principle, one way of calculating this number is counting all possible paths and obtain the associated probabilities. In most cases, this procedure is unpractical even for an OQW acting on  density matrices of low order in the $N$-path, as general products of $L$'s and $R$'s may be difficult to control. One situation for which counting paths is feasible is when we can obtain some pattern in the multiplications of $L$ and $R$. For instance, consider
\begin{equation}\label{hadacoin}
L=\frac{1}{\sqrt{2}}\begin{bmatrix} 1 & 1 \\ 0 & 0 \end{bmatrix},\;\;\; R=\frac{1}{\sqrt{2}}\begin{bmatrix} 0 & 0 \\ 1 & -1 \end{bmatrix}
\end{equation}
Note that $L^*L+R^*R=I$ and, even though $LR\neq RL$, it is not difficult to recognize a pattern for the trace of products of $L$ and $R$. Write a density matrix as
\begin{equation}
\rho=\begin{bmatrix} \rho_{11} & \rho_{12} \\ \ov{\rho_{12}} & \rho_{22} \end{bmatrix}=\frac{1}{2}\begin{bmatrix} 1+x_3 & x_1+ix_2 \\ x_1-ix_2 & 1-x_3\end{bmatrix},\;\;\;x_1^2+x_2^2+x_3^2\leq 1, \; x_i\in\mathbb{R}
\end{equation}
Then one can show that if $C=C_n\cdots C_1$, where each $C_i\in\{L,R\}$ then
\begin{equation}
Tr(C\rho C^*)=\left\{
\begin{array}{rl}
\frac{1}{2^n}(1+2Re(\rho_{12}))=\frac{1}{2^n}(1+x_1), & \text{ if } C_1=L,\\
\frac{1}{2^n}(1-2Re(\rho_{12}))=\frac{1}{2^n}(1-x_1), & \text{ if } C_1=R
\end{array} \right.
\end{equation}
A calculation shows that for the $3$-path, we have $k_{31}(\rho)=4+2x_1$ and in the $4$-path we have $k_{41}(\rho)=9+2x_1$. Because of this we conjecture that for every $k\geq 2$,
\begin{equation}\label{conj1}
k_{k1}(\rho)=(k-1)^2+2x_1
\end{equation}
but a rigorous proof of this fact is, up to our knowledge, unknown. In particular, we have a mean hitting time that may depend on the initial density matrix, but its presence (at least in the case of order 2 densities) does not produce significant changes from the classical case (especially with respect to large $k$). One could then be interested in different choices of $L$ and $R$. Note that (\ref{hadacoin}) is obtained from dividing the Hadamard matrix in two pieces, so one could try to divide other unitary maps in two pieces and attempt a similar calculation. One notes that we are still able to obtain a multiplication pattern, that is, if we let
\begin{equation}\label{gencoin}
L=\begin{bmatrix} x & y \\ 0 & 0 \end{bmatrix},\;\;\;R=\begin{bmatrix} 0 & 0 \\ z & w \end{bmatrix},\;\;\;L^*L+R^*R=I, \;\;\;\rho=\begin{bmatrix} \rho_{11} & \rho_{12} \\ \ov{\rho_{12}} & \rho_{22}\end{bmatrix},
\end{equation}
then we can show that for $n=1,2,\dots$,
\begin{equation}
Tr(L^n\rho L^{n*})=|x^{n-1}|^2\big[\rho_{11}|x|^2+2Re(\rho_{12}x\ov{y})+\rho_{22}|y|^2\big],\;\;\;Tr(R^n\rho R^{n*})=|w^{n-1}|^2\big[\rho_{11}|z|^2+2Re(\rho_{12}z\ov{w})+\rho_{22}|w|^2\big],
\end{equation}
and from this and similar calculations we obtain the appropriate multiplication patterns for the other possible products. We omit the details. We believe the problem of calculating the mean hitting time for the OQW on the $N$-path associated to the pair of matrices (\ref{gencoin}) is also an open question, as it includes conjecture (\ref{conj1}) as a particular case.

\medskip

As an additional comment on eq. (\ref{conj1}), the following is a natural question. Can one obtain matrices $L$, $R$ with $L^*L+R^*R=I$ such that they are fair in some sense and such that these produce a more significant gain in the mean hitting time? Perhaps this needs to involve matrices which are more complicated than the ones of the form (\ref{gencoin}) so, apparently, a path counting argument seems to be unfeasible.

\medskip

Another related question is to examine the problem above when we consider density matrices of any size. In principle, it is possible that every off-diagonal entry of the density contributes to the computations of hitting times, and the setting presented here allows one to quantify how these notions differ from the classical results and to compare behaviour and statistics of distinct OQWs.

\medskip

{\bf 2. Target times of OQWs.} Recall the classical interpretation of the Random Target Lemma, expression (\ref{rtl_classica}): to say that $\sum_j \pi_j E_iT_j$ is constant (i.e., does not depend on $i$) means that if a walker begins at some chosen vertex, then the average number of edges he needs to cross in order to reach a destination chosen randomly does not depend on where the walker begins. Practical applications, such as the strategy of a tourist visiting places in a city are natural. As for the target time of OQWs,
\begin{equation}
t_{\odot}(\rho):=\sum_i Tr((\hat{D}^{-1}\hat{N})_{ij}\rho),
\end{equation}
we note that we may have a different interpretation: if an open quantum walker begins at any chosen vertex, then the average number of edges he needs to cross in order to reach a node chosen randomly depends, in general, on the initial density matrix and this may imply (via the topology of the walk and transition matrices) a dependence on where the walker begins. That is, due to the density matrix degree of freedom, the situation may turn out to be quite different from the classical case. In case the initial density does not influence the target time (see e.g. Example \ref{carboneex}), this can be seen as the walk having a classical character, and a related question in this situation is to ask whether this happens to every ergodic OQW.

\medskip

As a consequence of the above, we may ask whether this generalized quantity has more information on the quantum dynamics studied in this work. In this line of work, we mention another possible application:

\medskip

{\bf 2.1. System-bath models.} In \cite{sinayskiy2}, Sinayskiy and Petruccione study the microscopic derivation of OQWs. The continuous-time evolution of the walk is studied, with iterative dynamics obtained via a discretization procedure. For instance, for an OQW on the $N$-path, a system and system-bath Hamiltonian are specified, and we can obtain the generalized master equations:
$$\frac{d}{dt}\rho_1(t)=\gamma(\langle n\rangle +1)S\rho_2(t)S-\frac{\gamma\langle n\rangle}{2}\{S^2,\rho_1(t)\}_+$$
\begin{equation}
\frac{d}{dt}\rho_N(t)=\gamma\langle n\rangle S\rho_{N-1}(t)S-\frac{\gamma(\langle n\rangle+1)}{2}\{S^2,\rho_N(t)\}_+
\end{equation}
$$\frac{d}{dt}\rho_i(t)=\gamma(\langle n\rangle +1)\Big(S\rho_{i-1}(t)S-\frac{1}{2}\{S^2,\rho_i(t)\}_+\Big) +\gamma\langle n\rangle\Big(S\rho_{i+1}(t)S-\frac{1}{2}\{S^2,\rho_i(t)\}_+\Big),\;\;\;i=2,\dots,N-1$$
Above, $\langle n\rangle$, and $\gamma$ are parameters, $S=S^*=\alpha\sigma_z+\beta I$, $\alpha,\beta\in\mathbb{R}$.
For the discretization, we replace the time derivative by the finite difference with a small time step $\Delta$, that is, $\rho_i'(t)\mapsto (\rho_i(t+\Delta)-\rho_i(t))/\Delta$. Then, one obtains the jump operators $B_{ij}$, which depend on $S$, $\gamma$, $\langle n\rangle$, $\Delta$, and the iterative evolution is now given by \cite{sinayskiy2}
$$\rho_1^{(n+1)}=B_{11}\rho_1^{(n)}B_{11}^*+B_{12}\rho_2^{(n)}B_{12}^*$$
\begin{equation}
\rho_N^{(n+1)}=B_{NN}\rho_N^{(n)}B_{NN}^*+B_{N,N-1}\rho_{N-1}^{(n)}B_{N,N-1}^*
\end{equation}
$$\rho_i^{(n+1)}=B_{ii}\rho_i B_{ii}^*+B_{i,i-1}\rho_{i-1} B_{i,i-1}^*+B_{i,i+1}\rho_{i+1} B_{i,i+1}^*,\;\;\;i=2,\dots,N-1$$
From the Random Target Lemma applied to this setting, we learn the following:
if the OQW begins at some chosen vertex, then the average number of edges the walker needs to cross in order to reach a node chosen randomly does not depend directly on the initial vertex, but depends on the initial density matrix, and this may be used to quantify the speed of the walk. From these observations, we also ask: a) what are the initial densities such that the associated target times are a maximum or minimum? b) How does this time varies with respect to the various parameters indicated above? These questions may also lead to comparisons with what happens in a setting of unitary walks.

\medskip

{\bf 3. Dissipative quantum computing algorithms.} Another area where OQWs have an important  application is dissipative quantum computing. As it is known, decoherence in certain amounts have been found to be useful in quantum computation \cite{kendon} and, in \cite{sinayskiy3}, it is  shown that the OQW implementation of the Toffoli gate and the Quantum Fourier Transform with 3 and 4 qubits outperforms the dissipative model discussed by Verstraete et al. \cite{verst}. A simple gate given by a unitary operator $U$ is shown in Figure 2, with matrices $B_{11}=\sqrt{\lambda}I$, $B_{22}=\sqrt{\omega}I$, $B_{21}=\sqrt{\omega}U$, $B_{12}=\sqrt{\lambda}U^*$ (as usual, $B_{ij}$ accounts for the transition from vertex $j$ to $i$).
\begin{center}
\begin{figure}[h]
\begin{tikzpicture}[->,>=stealth',shorten >=1pt,auto,node distance=2.8cm,
                    semithick]
  \tikzstyle{every state}=[fill=black,draw=none,text=white]

  \node[main node] (B)  {$1$};
  \node[main node] (E) [right of=B] {$2$};
  \path (B) edge   [loop above]     node {$\sqrt{\lambda}I$} (B)
        (B) edge   [bend left]     node {$\sqrt{\omega}U$} (E)
        (E) edge   [bend left]     node {$\sqrt{\lambda}U^*$} (B)
        (E) edge   [loop above]     node {$\sqrt{\omega}I$} (E);
\end{tikzpicture}
\caption{A gate given by unitary operator can be implemented by a 2-node OQW. The transition matrices depend on parameters $\omega$, $\lambda$, with $\lambda+\omega=1$, which have a physical meaning that can be understood in terms of the underlying microscopic model of the system \cite{sinayskiy2,sinayskiy3}. If the initial state of the quantum system $|\psi_1\rangle$ is prepared in the node $1$, then after performing the open quantum walk the system reaches the steady state $\rho_{SS}=\lambda|\psi_1\rangle\langle\psi_1|\otimes|1\rangle\langle 1|+\omega U|\psi_1\rangle\langle\psi_1|U^*\otimes|2\rangle\langle 2|$.
}
\end{figure}
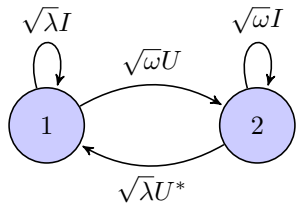
\end{center}
Then, the result of the gate application can be detected in node 2 with probability $\omega$.
 In \cite{sinayskiy3}, the OQW implementation of Toffoli gate (which needs 13 unitary operators) is examined, so that the number of steps needed to reach the steady state, and the probability of detection of the result of the computation in the steady state are obtained as a function of the walk parameter $\omega$. One may be interested in the average number of steps (i.e., the mean hitting time) and the dependence of this quantity in terms of the initial state. However, one basic point is to be noted: the calculation of hitting times for OQWs involves traces of the form $x_k=Tr(V_k\cdots V_1\rho V_1^*\cdots V_k^*)$, for $V_i$ matrices  associated to the transitions. If all matrices $V_i$ are multiples of unitary matrices, e.g., $V_i=\sqrt{c_i}U_i$, $c_i\in\mathbb{C}, U_i\in U(n)$, then by the linearity of the trace the computation of $x_k$ reduces trivially to the product  $x_k=c_1\cdots c_k$, and so this resembles a classical calculation where there is no dependence on the initial state. We may have a different situation, however, in the case that we have a circuit with nonunitary gates \cite{terashima}. A more detailed discussion on these matters will appear in a future work.

\medskip

{\bf 4. Other results associated to mean hitting times.} Besides the Mean Hitting Time Formula, quite often we make use of some expression which is a consequence of it, the Random Target Lemma being an important example. Generally speaking, it is well-known that we can relate mean hitting times and occupation times via a number of identities which consist of important tools in classical probability. We mention one example:
\begin{pro}\cite{aldous}
Consider a finite state, irreducible Markov chain with unique stationary measure $\pi$. If $i\neq j$, then
$$
E_i(\textrm{number of visits to j before time $T_i$})=\pi_j(E_j T_i+E_i T_j),
$$
where $T_i$ is the time of first visit to vertex $i$.
\end{pro}
See [\cite{aldous}, Ch. 2, Sec. 2.2] for other identities. Then, it is a natural question to ask whether we can construct a similar result in an open quantum setting. If a quantum version in fact exists, how similar is it to the classical result? As in the Random Target Lemma, we may have some different statistical interpretation due to the density matrix degree of freedom.

\medskip

{\bf Structure of this work.} In Section \ref{sec2} we review completely positive maps (quantum channels) and basic properties of OQWs, including calculation of probabilities in such setting. We discuss the block form, the matrix representation of OQWs and the relation between them. In Section \ref{sec3} we prove basic properties of the fundamental matrix for OQWs and in Section \ref{sec4} we make a remark on calculations of hitting probabilities. In Section \ref{sec5} we prove Theorem \ref{quantum_aldous3}. The structure of the proof of this theorem is inspired by \cite{aldous} and \cite{bremaud} and we also consider a numerical example. In Section \ref{sec77} we prove Corollary \ref{rtlemma} and in Section \ref{sec6} we prove Theorem \ref{chzhthm}.

\bigskip

\section{Open Quantum Random Walks and representations}\label{sec2}

We review basic facts about completely positive maps, more details of which can be seen for instance in \cite{bhatia}. Let $\Phi: M_n(\mathbb{C})\to M_n(\mathbb{C})$ be a linear map on the algebra of order $n$ matrices with complex coefficients. Recall that a matrix $A\in M_n(\mathbb{C})$ is positive semidefinite if $A^*=A$ and $\langle Av,v\rangle\geq 0$ for all $v\in\mathbb{C}^n$, and we denote this fact by $A\geq 0$. We say $\Phi$ is a {\bf positive} operator whenever $A\geq 0$ implies $\Phi(A)\geq 0$. Define for each $k\geq 1$, $\Phi_k: M_k(M_n(\mathbb{C}))\to M_k(M_n(\mathbb{C}))$,
\begin{equation}
\Phi_k(A)=\begin{bmatrix} \Phi(A_{11}) & \cdots & \Phi(A_{1k}) \\ \vdots & \cdots & \vdots \\ \Phi(A_{k1}) & \cdots & \Phi(A_{kk}) \end{bmatrix} ,\;\;\; A\in M_k(M_n(\mathbb{C})), \; A_{ij}\in M_n(\mathbb{C})
\end{equation}
We say $\Phi$ is {\bf $k$-positive} if $\Phi_k$ is positive, and we say $\Phi$ is {\bf completely positive (CP)} if $\Phi_k$ is positive for every $k=1, 2, 3,\dots$. We have that CP maps can be written in the Kraus form \cite{bhatia}: there are $B_i\in M_n(\mathbb{C})$ such that
\begin{equation}\label{krausform}
\Phi(\rho)=\sum_i B_i\rho B_i^*
\end{equation}
We say $\Phi$ is {\bf trace-preserving} if $Tr(\Phi(\rho))=Tr(\rho)$ for all $\rho\in M_n(\mathbb{C})$, which is equivalent to $\sum_i B_i^*B_i=I$, or $\Phi^*(I)=I$. We say $\Phi$ is {\bf unital} if $\Phi(I)=I$, which is equivalent to $\sum_i B_iB_i^*=I$. Trace-preserving completely positive (CPT) maps are also called {\bf quantum channels} in mathematical physics literature \cite{nielsen,petz}. Also recall that if $A\in M_n(\mathbb{C})$ there is the corresponding vector representation $vec(A)$ associated to it, given by stacking together the matrix rows. For instance, if $n=2$,
\begin{equation}
A=\begin{bmatrix} a_{11} & a_{12} \\ a_{21} & a_{22}\end{bmatrix}\;\;\;\Longrightarrow \;\;\; vec(A):=\begin{bmatrix} a_{11} \\ a_{12} \\ a_{21} \\ a_{22}\end{bmatrix}
\end{equation}
The $vec$ mapping satisfies $vec(AXB^T)=(A\otimes B)vec(X)$ for any $A, B, X$ square matrices (see \cite{hj2}, noting that in such reference $vec$ means stacking the matrix columns instead of rows) so in particular, $vec(BXB^*)=vec(BX\ov{B}^T)=(B\otimes \ov{B})vec(X)$,
from which we can obtain the {\bf matrix representation} $[\Phi]$ for the CP map (\ref{krausform}):
\begin{equation}\label{matrep}
[\Phi]:=\sum_{i} B_{i}\otimes \ov{B_{i}},
\end{equation}
and given any matrix $V\in M_n(\mathbb{C})$ we define $[V]:=V\otimes\ov{V}\in M_{n^2}(\mathbb{C})$. We recall the well-known fact that the matrix representation of a CPT map $\Phi: M_n(\mathbb{C})\to M_n(\mathbb{C})$ is independent of the Kraus representation considered. The proof of this result is a simple consequence of the unitary equivalence of Kraus matrices for a given quantum channel \cite{petz}.

\medskip

\subsection{Open Quantum Random Walks (OQWs)} We denote a vector in a $n$-dimensional complex vector space $V$ by its $n$ complex coordinates, or by the corresponding vector column, and the dual operator by its conjugate transpose, that is,
\begin{equation}
v=(v_1,\dots,v_n)\in V\;\;\; |v\rangle:=\begin{bmatrix} v_1 \\ \vdots \\ v_n\end{bmatrix},\;\;\;\langle v|:=\begin{bmatrix} \ov{v_1} & \cdots & \ov{v_n}\end{bmatrix}
\end{equation}
With this notation, $|v \rangle \langle v|$ denotes the projection map on the direction of $v$. Let $\mathcal{K}$ denote a separable Hilbert space and let $\{|i\rangle\}_{i\in \mathbb{Z}}$ be an orthonormal basis for such space (indexed by the integers in case $\mathcal{K}$ is infinite dimensional). The elements of such basis will be called {\bf sites} (the vertices of a given graph) and denoted by $|i\rangle$, or just $i$ for simplicity. Let $\mathcal{H}$ be another Hilbert space, which will describe the {\bf degrees of freedom} given at each point of $\mathbb{Z}$. Then we will consider the space $\mathcal{H}\otimes\mathcal{K}$. For each pair $i,j$ we associate a bounded operator $B_{ij}$ on $\mathcal{H}$. This operator describes the effect of passing from $|j\rangle$ to $|i\rangle$. We will assume that for each $j$, $\sum_i B_{ij}^{*}B_{ij}=I$, where, if infinite, such series is strongly convergent. This constraint means that the sum of all the effects leaving the site $j$ is $I$.

\medskip

We will consider density matrices on $\mathcal{H}\otimes\mathcal{K}$ with the particular form given by
(\ref{dens_attal}). For a given initial state of such form, the Open Quantum Random Walk induced by the $B_{ij}$ is given by (\ref{oqrw_def}). We note that by defining $M_{ij}=B_{ij}\otimes|i\rangle\langle j|$ we have a Kraus representation for $\Phi$, given by
\begin{equation}
\Phi(\rho)=\sum_{ij} M_{ij}\rho M_{ij}^*
\end{equation}
We say the OQW is {\bf finite} if $\dim(\mathcal{H}\otimes\mathcal{K})<\infty$, that is, if the individual density matrices are finite dimensional and if the walk acts on a finite number of sites.
In this work we will assume that $\mathcal{H}=\mathbb{C}^n$ \and $\mathcal{K}=\mathbb{C}^k$.

\medskip

\begin{remark}\label{oqsremark} (Calculation of probabilities). Following \cite{attal}, we have a statistical interpretation of the traces of the matrices associated to each site: if $\rho^{(0)}=\sum_i \rho_i^{(0)}\otimes |i\rangle\langle i|=(\rho_1^{(0)},\dots, \rho_n^{(0)})$ then
\begin{equation}
\rho^{(r)}=\Phi^r(\rho^{(0)})=\sum_i \rho_i^{(r)}\otimes |i\rangle\langle i|,
\end{equation}
and the probability of occurrence of the open quantum walk at site $|i\rangle$ at time $r$ equals $Tr(\rho_i^{(r)})$. For instance, if we have a nearest neighbor interaction on $\mathbb{Z}$,
\begin{equation}
\Phi(\rho)=\sum_{i\in\mathbb{Z}}[R\rho_{i-1}R^*+L\rho_{i+1}L]\otimes|i\rangle\langle i|
\end{equation}
and $\rho^{(0)}=\rho_0\otimes |0\rangle\langle 0|$ then $\rho^{(1)}=L\rho_0 L^*\otimes|-1\rangle\langle -1|+R\rho_0 R^*\otimes|1\rangle\langle 1|$ and
\begin{equation}
\rho^{(2)}=L^2\rho_0 L^{2*}\otimes|-2\rangle\langle -2|+(LR\rho_0 R^*L^*+RL\rho_0 L^*R^*)\otimes|0\rangle\langle 0|+R^2\rho_0 R^{2*}\otimes|2\rangle\langle 2|,
\end{equation}
so the probability of reaching sites $|-2\rangle$, $|0\rangle$  and $|2\rangle$ in two steps, given that the walk started at site $|0\rangle$ with initial density $\rho_0$ is given by $Tr(L^2\rho_0 L^{2*})$, $Tr(LR\rho_0 R^*L^*+RL\rho_0 L^*R^*)$  and $Tr(R^2\rho_0 R^{2*})$, respectively. This is one of the simplest examples of OQWs. If $i$ is allowed to vary in $\mathbb{Z}$, we have a nearest neighbor OQW on the integer line, with left and right transitions given by matrices $L$ and $R$, respectively, these satisfying $L^*L+R^*R=I$. It is clear that one may consider site dependent transitions and also more than just nearest neighbor interactions. This setting is also suitable to consider open quantum walks on graphs.
\end{remark}

\subsection{Block form of OQWs}

We recall that if $\Phi$ is an OQW, let $\hat{\Phi}$ denote its {\bf block representation}: for 2 sites,
\begin{equation}\label{oqw2sites}
\Phi(\rho)=(B_{11}\rho_1B_{11}^{*}+B_{12}\rho_2B_{12}^{*})\otimes |1\rangle\langle 1|+(B_{21}\rho_1B_{21}^{*}+B_{22}\rho_2B_{22}^{*})\otimes |2\rangle\langle 2|,
\end{equation}
we write
\begin{equation}
\hat{\Phi}=\begin{bmatrix} \hat{\Phi}_{11} & \hat{\Phi}_{12} \\ \hat{\Phi}_{21} & \hat{\Phi}_{22}\end{bmatrix}:=\begin{bmatrix} M_{B_{11}} & M_{B_{12}} \\ M_{B_{21}} & M_{B_{22}}\end{bmatrix},
\end{equation}
recall eq. (\ref{defconjug}). We give the analogous definition for more than 2 sites. Clearly a density $\rho=\rho_1\otimes|1\rangle\langle 1|+\rho_2\otimes|2\rangle\langle 2|$, $\rho_1, \rho_2\in M_2(\mathbb{C})$, can be identified with $[\rho_1 \;\;\rho_2]^T$. Then
we write for one iteration
\begin{equation}
\hat{\Phi}(\rho):=\begin{bmatrix} M_{B_{11}} & M_{B_{12}} \\ M_{B_{21}} & M_{B_{22}}\end{bmatrix}\begin{bmatrix} \rho_1 \\ \rho_2\end{bmatrix}=\begin{bmatrix} M_{B_{11}}\rho_1+M_{B_{12}}\rho_2\\ M_{B_{21}}\rho_1+M_{B_{22}}\rho_2\end{bmatrix}
\end{equation}
We also have
$$\hat{\Phi}(\hat{\Phi}(\rho))=\begin{bmatrix} M_{B_{11}}\Big(M_{B_{11}}\rho_1+M_{B_{12}}\rho_2\Big)+M_{B_{12}}\Big(M_{B_{21}}\rho_1+M_{B_{22}}\rho_2\Big)\\
M_{B_{21}}\Big(M_{B_{11}}\rho_1+M_{B_{12}}\rho_2\Big)+M_{B_{22}}\Big(M_{B_{21}}\rho_1+M_{B_{22}}\rho_2\Big)
\end{bmatrix}$$
\begin{equation}
=\begin{bmatrix} M_{B_{11}}^2+M_{B_{12}}M_{B_{21}} & M_{B_{11}}M_{B_{12}}+M_{B_{12}}M_{B_{22}} \\ M_{B_{21}}M_{B_{11}}+M_{B_{22}}M_{B_{21}} & M_{B_{21}}M_{B_{12}}+M_{B_{22}}^2\end{bmatrix}\begin{bmatrix} \rho_1\\ \rho_2\end{bmatrix}=\hat{\Phi}^2(\rho)
\end{equation}
We emphasize that in our notation the indices appearing in expressions are meant to be read from right to left. For instance, in the case of a nearest neighbor walk on $\mathbb{Z}$, $M_{B_{21}}M_{B_{12}}+M_{B_{22}}^2$ describes all possible ways of moving from site 2 to 2 in 2 steps.

\subsection{Matrix and vector representations}\label{generalizing_m}

If $\Phi$ is an OQW, let $[\Phi]$ denote its {\bf matrix representation}: for an OQW on 2 sites as in (\ref{oqw2sites}), we write
\begin{equation}\label{block_order2}
[\Phi]=\begin{bmatrix} [\hat{\Phi}_{11}] & [\hat{\Phi}_{12}] \\ [\hat{\Phi}_{21}] & [\hat{\Phi}_{22}]\end{bmatrix}:=\begin{bmatrix} [B_{11}] & [B_{12}] \\ [B_{21}] & [B_{22}]\end{bmatrix}=\begin{bmatrix} B_{11}\otimes \ov{B_{11}} & B_{12}\otimes \ov{B_{12}} \\ B_{21}\otimes \ov{B_{21}} & B_{22}\otimes \ov{B_{22}} \end{bmatrix},
\end{equation}
and we give the analogous definition for more than 2 sites. Recall $B_{ij}$ is the effect of passing from $|j\rangle$ to $|i\rangle$. We will make use of matrix representations together with block forms defined above. Expression (\ref{block_order2}) generalizes to $k$ sites in the natural way.

\begin{remark}
Given an OQW $\Phi$ we note that there is a clear identification between its block form $\hat{\Phi}$ and its matrix representation $[\Phi]$ and we may consider any of them whenever convenient. In particular, when we talk of convergence of a sequence of OQWs, we mean the entrywise matrix convergence of $[\Phi]$.
\end{remark}

Let us make an observation on the density matrices being considered. Suppose $\Phi$ is a 2-site OQW with transitions given by order 2 matrices, $B_{ij}\in M_2(\mathbb{C})$. Note that a density $\rho=\rho_1\otimes|1\rangle\langle 1|+\rho_2\otimes|2\rangle\langle 2|$, $\rho_1, \rho_2\in M_2(\mathbb{C})$, can be seen simply as a direct sum of effects $\rho=(\rho_1,\rho_2)$, which in turn may identified with $[vec(\rho_1) \;\;vec(\rho_2)]^T$  and this corresponds to a vector in $\mathbb{C}^8$. Also $[B_{ij}]\in M_4(\mathbb{C})$, $[\Phi]$ will be an order 8 matrix acting on two positive matrices $\rho_1, \rho_2\in M_2(\mathbb{C})$ and we write for one iteration, via the usual block matrix multiplication,
\begin{equation}
[\Phi](\rho):=\begin{bmatrix} [B_{11}] & [B_{12}] \\ [B_{21}] & [B_{22}]\end{bmatrix}\begin{bmatrix} vec(\rho_1) \\ vec(\rho_2)\end{bmatrix}=\begin{bmatrix} [B_{11}]vec(\rho_1)+[B_{12}]vec(\rho_2)\\ [B_{21}]vec(\rho_1)+[B_{22}]vec(\rho_2)\end{bmatrix}
\end{equation}
That is, corresponding to the matrix calculation $\rho\mapsto\sum_{ij} B_{ij}\rho B_{ij}^*$ we have the matrix-vector calculation $\hat{\Phi}\;vec(\rho)$, where $vec(\rho)=[vec(\rho_1) \;\;vec(\rho_2)]^T$. This is just the algebraic description of having a linear map (matrix) acting on an element of the vector space (i.e., a column vector).  Due to the correspondence between usual matrix multiplication and block matrix multiplication, we have that $vec^{-1}(\hat{\Phi}(vec(\rho)))$ gives the desired OQW iteration in direct sum of matrices once again. It is not difficult to show that the probability of reaching site $|j\rangle$ at time $r$, given that the walk has started at site $|i\rangle$, with initial density matrix $\rho_i$ equals
\begin{equation}\label{vecprob1}
P_{i,j;\rho}(r)=Tr(vec^{-1}[\Phi_{ji}^r]vec(\rho_i))
\end{equation}
Clearly, the price to pay for avoiding left and right multiplications of the form $V\rho V^*$ is the use of the $vec$ representation and then its inverse. For the sake of simplicity, formula (\ref{vecprob1}) will not used in this work and we will prefer to use the simpler expressions with the conjugation maps $M_{B_i}$. Nevertheless, we will occasionally show the explicit form of $[\Phi]$ for certain examples, with the purpose of gaining some calculational intuition.

\subsection{Example: PQ-matrices} An order $n$ matrix is a {\bf PQ-matrix} if it is a permutation of some diagonal matrix. Denote by $PQ_n$ the set of such matrices in $M_n(\mathbb{C})$. For instance, $PQ_2$ consists of all matrices which are diagonal ($a_{ij}=0$ if $i\neq j$) or antidiagonal ($a_{ii}=0$ for all $i$). Of course, for $n\geq 3$ the set $PQ_n$ allows for many other possibilities. One property possessed by PQ-matrices is the fact that if $V_1\cdots V_k$ is any product of PQ-matrices and $X$ is any matrix then the expressions $Tr(VX V^*)$ do not depend on nondiagonal entries of $X$, this being of computational convenience. Despite its simplicity, such objects may be used to describe dynamics which cannot be performed by classical Markov chains.
\begin{remark}

In terms of CP maps, one may consider PQ-quantum channels, i.e., a channel which admits a Kraus decomposition where each element is a PQ-matrix \cite{ls2014}. Clearly not every quantum channel is of this kind. These are among the simplest quantum channels and there are many examples in the literature, we refer the reader to \cite{ls2014} for more information.
\end{remark}

\begin{example} A basic class of examples consist of OQWs acting on $2$ sites, where the action on the densities occur via order 2 PQ-matrices. For instance, let $\{A,B\}$ be order $2$ PQ-matrices assuming that $A=(a_{ij})$ is diagonal and $B=(b_{ij})$ is antidiagonal. This implies a particular form for $[A]$ and $[B]$. If $A^*A+B^*B=I$ we may define, for instance, the following OQW:
\begin{equation}\label{a_pqex}
[\Phi]=\begin{bmatrix} [A] & [B] \\ [B] & [A] \end{bmatrix}=\begin{bmatrix} |a_{11}|^2 & 0 & 0 & 0 & 0 & 0 & 0 & |b_{12}|^2 \\ 0 & a_{11}\ov{a_{22}} & 0 & 0 & 0 & 0 & b_{12}\ov{b_{21}} & 0 \\ 0 & 0 & a_{22}\ov{a_{11}} & 0 & 0 & b_{21}\ov{b_{12}} & 0 & 0 \\ 0 & 0 & 0 & |a_{22}|^2 & |b_{21}|^2 & 0 & 0 & 0 \\ 0 & 0 & 0 & |b_{12}|^2 & |a_{11}|^2 & 0 & 0 & 0 \\ 0 & 0 & b_{12}\ov{b_{21}} & 0 & 0 & a_{11}\ov{a_{22}} & 0 & 0    \\ 0 & b_{21}\ov{b_{12}} & 0 & 0 & 0 & 0 & a_{22}\ov{a_{11}} & 0    \\ |b_{21}|^2 & 0 & 0 & 0 & 0 & 0 & 0 & |a_{22}|^2   \end{bmatrix}
\end{equation}
In this basic case, the two matrix rows consist of the same matrices (in a different order), but one can easily construct examples where all matrices are different (see Example \ref{numexample}).
\end{example}

\qee

\section{Fundamental matrix for ergodic OQWs}\label{sec3}

As in the previous sections, $k$ denotes the number of sites and $n$ is the degree of freedom at each site.  Let ${\Omega}$ denote the map such that each block of its matrix representation equals $[\hat{\Omega}_{pq}]=\frac{1}{kn}\sum_{i,j=1}^kE_{ij}\otimes E_{ij}$, for $p,q=1,\dots,k$ and $E_{ij}\in M_n(\mathbb{C})$ are the matrix units: $(E_{ij})_{rs}=1$ if $(r,s)=(i,j)$, and equals zero otherwise. For instance, if $k=n=2$, we have
\begin{equation}
[{\Omega}]=\begin{bmatrix} [\hat{{\Omega}}_{11}] & [\hat{{\Omega}}_{12}] \\ [\hat{{\Omega}}_{21}] & [\hat{{\Omega}}_{22}]\end{bmatrix}=\frac{1}{4}\begin{bmatrix} 1 & 0 & 0 & 1 & 1 & 0 & 0 & 1 \\ 0 & 0 & 0 & 0 & 0 & 0 & 0 & 0 \\0 & 0 & 0 & 0 & 0 & 0 & 0 & 0 \\ 1 & 0 & 0 & 1 & 1 & 0 & 0 & 1 \\ 1 & 0 & 0 & 1 & 1 & 0 & 0 & 1 \\0 & 0 & 0 & 0 & 0 & 0 & 0 & 0 \\0 & 0 & 0 & 0 & 0 & 0 & 0 & 0 \\ 1 & 0 & 0 & 1 & 1 & 0 & 0 & 1 \end{bmatrix}
\end{equation}
Note that $\hat{\Omega}^2=\hat{\Omega}$. The idea behind this definition is that there are OQWs such that the iterates of its block representation converge to $\hat{\Omega}$. This should be compared with the fact that the columns (rows) of a stochastic matrix corresponding to finite irreducible aperiodic Markov chains converge to the unique stationary probability vector. Note that $\hat{\Omega}$ has as fixed point the density
\begin{equation}
\rho=\frac{1}{kn}\sum_{i=1}^k I_n\otimes|i\rangle\langle i|=\frac{1}{kn}[I_n\;\dots\;I_n]^T
\end{equation}
To see this in the case $n=k=2$, just note that $\rho=\frac{1}{4}[I_2\; I_2]$ and the vector representation for $\rho$ equals
\begin{equation}
vec(\rho)=\frac{1}{4}\begin{bmatrix} 1 & 0 & 0 & 1 & 1 & 0 & 0 & 1 \end{bmatrix}^T
\end{equation}
so we clearly have $\hat{\Omega}(vec(\rho))=vec(\rho)$. Another basic fact is the following, which will be used shortly:
\begin{equation}\label{basicbrem1}
\hat{\Phi}\hat{\Omega}=\hat{\Omega}\hat{\Phi}=\hat{\Omega}
\end{equation}

Now we make a calculation. Let $\Phi\in\mathcal{E}$ and note that
\begin{equation}\label{z_aux1}
(\hat{\Phi}-\hat{\Omega})^r=\sum_{k=0}^r {r\choose k} (-1)^{r-k}\hat{\Phi}^k\hat{\Omega}^{r-k}=\hat{\Phi}^r+\sum_{k=0}^{r-1}{r\choose k}(-1)^{r-k}\hat{\Omega}=\hat{\Phi}^r-\hat{\Omega},
\end{equation}
since $\sum_{k=0}^{r}{r\choose k}(-1)^{r-k}=\sum_{k=0}^{r-1}{r\choose k}(-1)^{r-k}+1=(a+b)^r=0$ if $a=1$, $b=-1$. Now let $\hat{A}:=\hat{\Phi}-\hat{\Omega}$. Then if $\hat{I}$ is the identity block matrix,
\begin{equation}\label{block_id}
[\hat{I}]=\begin{bmatrix} [I_n] & 0 & \cdots & 0 \\ 0 & [I_n] & \cdots & 0 \\ \vdots & \vdots & \ddots  & \vdots \\ 0 & \cdots & \cdots & [I_n]\end{bmatrix},\;\;\;\hat{I}\in M_{kn^2}(\mathbb{C}),
\end{equation}
then
\begin{equation}
(\hat{I}-\hat{A})(\hat{I}+\hat{A}+\hat{A}^2+\cdots+\hat{A}^{r-1})=\hat{I}-\hat{A}^r=\hat{I}-(\hat{\Phi}-\hat{\Omega})^r=\hat{I}-\hat{\Phi}^r+\hat{\Omega}
\end{equation}
If $r\to\infty$ then
\begin{equation}
(\hat{I}-\hat{A})(\hat{I}+\sum_{r\geq 1}\hat{A}^r)=\hat{I}
\end{equation}
This shows that $\hat{I}-\hat{A}=\hat{I}-\hat{\Phi}+\hat{\Omega}$ is invertible, with inverse
\begin{equation}
\hat{I}+\sum_{r\geq 1}(\hat{\Phi}-\hat{\Omega})^r=\hat{I}+\sum_{r\geq 1}(\hat{\Phi}^r-\hat{\Omega})
\end{equation}
This allows us to define the fundamental matrix of ergodic OQWs as described in the introduction:  for every $\Phi\in\mathcal{E}$ acting on a finite collection of sites, define
\begin{equation}
\hat{\mathcal{Z}}:=\hat{I}+\sum_{r\geq 1}(\hat{\Phi}^r-\hat{\Omega})=(\hat{I}-\hat{\Phi}+\hat{\Omega})^{-1}
\end{equation}
The following will be needed later in this work.

\begin{lemma}\label{cflema1}
Let $\Phi\in\mathcal{E}$. Then a) $\hat{\mathcal{Z}}\hat{\Omega}=\hat{\Omega}\hat{\mathcal{Z}}=\hat{\Omega}$. b) $\hat{\mathcal{Z}}(I-\hat{\Phi})=\hat{I}-\hat{\mathcal{Z}}\hat{\Omega}=\hat{I}-\hat{\Omega}$. c) $(\hat{I}-\hat{\Phi})\hat{\mathcal{Z}}=\hat{I}-\hat{\Omega}\hat{\mathcal{Z}}=\hat{I}-\hat{\Omega}$
\end{lemma}
{\bf Proof.} a)  Write $\hat{\mathcal{Z}}(\hat{I}-\hat{\Phi}+\hat{\Omega})=\hat{I}$ so $\hat{\mathcal{Z}}-\hat{\mathcal{Z}}\hat{\Phi}+\hat{\mathcal{Z}}\hat{\Omega}=\hat{I}$, with $\hat{I}$ given by (\ref{block_id}). Right multiply by  $\hat{\Omega}$ on both sides, so
\begin{equation}
\hat{\mathcal{Z}}\hat{\Omega}-\hat{\mathcal{Z}}\hat{\Phi}\hat{\Omega}+\hat{\mathcal{Z}}\hat{\Omega}\hat{\Omega}=I\hat{\Omega}
\end{equation}
By (\ref{basicbrem1}), we get
\begin{equation}
\hat{\mathcal{Z}}\hat{\Omega}-\hat{\mathcal{Z}}\hat{\Omega}+\hat{\mathcal{Z}}\hat{\Omega}=\hat{\Omega}\;\Longrightarrow\; \hat{\mathcal{Z}}\hat{\Omega}=\hat{\Omega}
\end{equation}
The proof of the other equality is similar.
b) Just note that $(\hat{I}-\hat{\Phi}+\hat{\Omega})\hat{\mathcal{Z}}=\hat{I}$, so $(\hat{I}-\hat{\Phi})\hat{\mathcal{Z}}=\hat{I}-\hat{\Omega}\hat{\mathcal{Z}}=\hat{I}-\hat{\Omega}$, the proof of the other equality being similar. The proof of c) is similar to b).
\qed

\section{A remark on the probability of first visit}\label{sec4}

Let $\Phi$ be an OQW and recall the matrix of operations $\hat{H}$. We claim that
\begin{equation}
\hat{H}\hat{\Phi}\begin{bmatrix} \rho_1 \\ 0 \end{bmatrix}=\hat{H}\begin{bmatrix} \rho_1 \\ 0 \end{bmatrix}
\end{equation}
In fact, recalling $B_{ij}$ is the effect of passing from $|j\rangle$ to $|i\rangle$ we have
$$\hat{H}\hat{\Phi}\begin{bmatrix} \rho_1 \\ 0 \end{bmatrix}=\begin{bmatrix} \sum_{C\in\pi(1\leftarrow 1)}M_C & \sum_{C\in\pi(1\leftarrow 2)}M_C\\
\sum_{C\in\pi(2\leftarrow 1)}M_C & \sum_{C\in\pi(2\leftarrow 2)}M_C\end{bmatrix}\begin{bmatrix} M_{B_{11}}\rho_1 \\ M_{B_{21}}\rho_1 \end{bmatrix}$$
\begin{equation}
=\begin{bmatrix} \sum_{C\in\pi(1\leftarrow 1)}M_CM_{B_{11}}\rho_1+\sum_{C\in\pi(1\leftarrow 2)}M_CM_{B_{21}}\rho_1\\ \sum_{C\in\pi(2\leftarrow 1)}M_CM_{B_{11}}\rho_1+\sum_{C\in\pi(2\leftarrow 2)}M_CM_{B_{21}}\rho_1 \end{bmatrix}=\begin{bmatrix} \sum_{C\in\pi(1\leftarrow 1)}M_C\rho_1 \\ \sum_{C\in\pi(2\leftarrow 1)}M_C\rho_1\end{bmatrix}=\hat{H}vec\Big(\begin{bmatrix} \rho_1 \\ 0 \end{bmatrix}\Big)
\end{equation}
And analogously for $[0 \; \rho_2]^T$ and so we obtain
\begin{equation}\label{genopeq1}
\sum_{C\in\pi(j\leftarrow i)}M_C\rho_i=\sum_{C\in\pi(j\leftarrow 1)}M_CM_{B_{1i}}\rho_i+\sum_{C\in\pi(j\leftarrow 2)}M_CM_{B_{2i}}\rho_i,\;\;\;i,j=1,2
\end{equation}
The above resoning holds for $k>2$ sites. We note again that in OQWs, when we specify that a walk is located at a certain site $|j\rangle$, one should also indicate the matrix degree of freedom on that site. We denote this by $\rho_j\otimes|j\rangle\langle j|$ (see expression (\ref{dens_attal})).

\medskip

 Note that we can also write
\begin{equation}\label{OQW_version1}
h_{ji}(\rho_i)=\sum_k h_{jk}(B_{ki}\rho_i B_{ki}^*)=\sum_k Tr(B_{ki}\rho_i B_{ki}^*)h_{jk}\Big(\frac{B_{ki}\rho_i B_{ki}^*}{Tr(B_{ki}\rho_j B_{ki}^*)}\Big),
\end{equation}
so a classical expression is recovered when we take order 1 density matrices thus eliminating the matrix dependence of $h_{ji}$ for any given site $i$: in this particular case we have $h_{ji}(\rho_i)=h_{ji}$, for $B_{ki}=\sqrt{p_{ki}}I$ we get $Tr(B_{ki}\rho_i B_{ki}^*)=p_{ki}$, and (\ref{OQW_version1}) becomes
\begin{equation}
\sum_k h_{jk}p_{ki}=h_{ji},\;\;\;i\neq j
\end{equation}
and $h_{ii}=1$. Note that the above equation is just the well-known matrix equation $hP=h$ from the classical Markov chain theory \cite{grimmett}.

\section{Proof of Theorem \ref{quantum_aldous3}}\label{sec5}

First we make a simple adaptation from a classical probability reasoning. First, note that $k_{ii}(\rho)=0$ for every $\rho$ density and for all $i$. Also, define
\begin{equation}
k_{ij}(\rho_j|X_1=l):=1+k_{il}\Big(\frac{B_{lj}\rho_j B_{lj}^*}{Tr(B_{lj}\rho_j B_{lj}^*)}\Big)
\end{equation}
Then if $i\neq j$,
$$k_{ij}(\rho_j)=\sum_l k_{ij}(\rho_j|X_1=l)Tr(B_{lj}\rho_j B_{lj}^*)=\sum_l\Big[ 1+k_{il}\Big(\frac{B_{lj}\rho_j B_{lj}^*}{Tr(B_{lj}\rho_j B_{lj}^*)}\Big)\Big]Tr(B_{lj}\rho_j B_{lj}^*)$$
$$=1+k_{ii}\Big(\frac{B_{ij}\rho_j B_{ij}^*}{Tr(B_{ij}\rho_j B_{ij}^*)}\Big)Tr(B_{ij}\rho_j B_{ij}^*)+\sum_{l\neq i}k_{il}\Big(\frac{B_{lj}\rho_j B_{lj}^*}{Tr(B_{lj}\rho_j B_{lj}^*)}\Big)Tr(B_{lj}\rho_j B_{lj}^*)$$
\begin{equation}\label{t_onk_exp}
=1+\sum_{l\neq i}k_{il}\Big(\frac{B_{lj}\rho_j B_{lj}^*}{Tr(B_{lj}\rho_j B_{lj}^*)}\Big)Tr(B_{lj}\rho_j B_{lj}^*)
\end{equation}
We note that in (\ref{t_onk_exp}), a trace term is introduced to emphasize that $k_{ij}$ acts on density matrices. A similar term is also used to recover a classical probability expression in (\ref{OQW_version1}), regarding the hitting probability. Let us write (\ref{t_onk_exp}) without the trace terms:
\begin{equation}
k_{ij}(\rho_j)=1+\sum_{l\neq i} k_{il}(B_{lj}\rho_j B_{lj}^*) \;\Longrightarrow \; c=k_{ij}(c\rho_j)-\sum_{l\neq i} k_{il}(B_{lj}c\rho_j B_{lj}^*),\;\;\;\forall\;c\in\mathbb{R}
\end{equation}
Let $\hat{D}=\hat{D}(\hat{K}):=diag(\hat{k}_{11},\dots,\hat{k}_{nn})$ and $\hat{L}:=\hat{K}-(\hat{K}-\hat{D})\hat{\Phi}$. By considering densities of the form $[0 \; \cdots \; \rho_j\; 0 \;\cdots \;0]^T$ where $\rho_j$ appears on the $j$-th position, we can write
$\hat{L}(c\rho)=\hat{K}(c\rho)-(\hat{K}-\hat{D})\hat{\Phi}(c\rho)$ and this implies, for all $i$ and all $c\in\mathbb{R}$,
\begin{equation}
Tr(\hat{L}_{ij}(c\rho_j))=Tr(\hat{K}_{ij}(c\rho_j)-[(\hat{K}-\hat{D})\hat{\Phi}]_{ij}(c\rho_j))=k_{ij}(c\rho_j)-\sum_{l\neq i} k_{il}(M_{B_{lj}}c\rho_j)=c
\end{equation}
For instance, in the case $k=3$, the summation on $l$ above arises from the product
\begin{equation}
(\hat{K}-\hat{D})\hat{\Phi}=\begin{bmatrix} 0 & \hat{k}_{12} & \hat{k}_{13} \\ \hat{k}_{21} & 0 & \hat{k}_{23} \\ \hat{k}_{31} & \hat{k}_{32} & 0 \end{bmatrix}\begin{bmatrix} \hat{\Phi}_{11} & \hat{\Phi}_{12} & \hat{\Phi}_{13} \\ \hat{\Phi}_{21} & \hat{\Phi}_{22} & \hat{\Phi}_{23} \\ \hat{\Phi}_{31} & \hat{\Phi}_{32} & \hat{\Phi}_{33}\end{bmatrix}
=\begin{bmatrix}
\hat{k}_{12}\hat{\Phi}_{21}+\hat{k}_{13}\hat{\Phi}_{31} & \hat{k}_{12}\hat{\Phi}_{22}+\hat{k}_{13}\hat{\Phi}_{32} &
\hat{k}_{12}\hat{\Phi}_{23}+\hat{k}_{13}\hat{\Phi}_{33} \\

\hat{k}_{21}\hat{\Phi}_{11}+\hat{k}_{23}\hat{\Phi}_{31} & \hat{k}_{21}\hat{\Phi}_{12}+\hat{k}_{23}\hat{\Phi}_{32} &
\hat{k}_{21}\hat{\Phi}_{13}+\hat{k}_{23}\hat{\Phi}_{33} \\

\hat{k}_{31}\hat{\Phi}_{11}+\hat{k}_{32}\hat{\Phi}_{21} & \hat{k}_{31}\hat{\Phi}_{12}+\hat{k}_{32}\hat{\Phi}_{22} &
\hat{k}_{31}\hat{\Phi}_{13}+\hat{k}_{32}\hat{\Phi}_{23}
  \end{bmatrix}
\end{equation}
We have concluded:
\begin{lemma}\label{very_usedlater}
Let $\Phi$ denote a finite ergodic OQW and let $\hat{\mathcal{Z}}$ denote its fundamental matrix. Let $\hat{K}$ be given as above. Let
$\hat{D}=\hat{D}=diag(\hat{k}_{11},\dots,\hat{k}_{nn})$ and let $\hat{L}:=\hat{K}-(\hat{K}-\hat{D})\hat{\Phi}$. Then if $\rho_j$ is a density matrix concentrated on site $j$ then for all $i$ and all $c\in\mathbb{R}$, $Tr(\hat{L}_{ij}c\rho_j)=Tr(c\rho_j)=c$.
\end{lemma}

From the lemma above we have, for instance, $Tr(\hat{L}_{12}\rho)=Tr(\rho)$ for $\rho$ density matrix on site $2$. This kind of calculation can be used to recover the classical case. In fact, if $\rho$ is a density matrix then for an OQW acting on 2 sites,
\begin{equation}
1=Tr(\hat{k}_{11}\rho)-Tr(\hat{k}_{12}M_{B_{21}}\rho)\;\Longrightarrow \; k_{11}(\rho)=1+k_{12}(M_{B_{21}}\rho),
\end{equation}
\begin{equation}
1=Tr(\hat{k}_{12}\rho)-Tr(\hat{k}_{12}M_{B_{22}}\rho)\;\Longrightarrow \; k_{12}(\rho)=1+k_{12}(M_{B_{22}}\rho)
\end{equation}
That is, we have obtained a density dependent version of the classical minimality theorem for mean hitting times. If we choose the $B_{ij}$ as multiples of the identity then the above equations become particular expressions of the result for mean hitting times for classical Markov chains \cite{grimmett,ls2015}:
\begin{equation}
k_{ij}=1+\sum_{l:l\neq i}k_{il}p_{lj},\;\;\;i\neq j
\end{equation}

\begin{lemma}
\label{quantum_aldous2}
Let $\Phi$ denote a finite ergodic OQW and let $\hat{\mathcal{Z}}$ denote its fundamental matrix. Let $\hat{K}$ be given by (\ref{k_eq1}). Let
$\hat{D}:=diag(\hat{k}_{11},\dots,\hat{k}_{nn})$, $\hat{L}:=\hat{K}-(\hat{K}-\hat{D})\hat{\Phi}$ and $\hat{N}:=\hat{K}-\hat{D}$. Then
\begin{equation}
\hat{N}_{ij}=(\hat{D}\hat{\mathcal{Z}})_{ii}-(\hat{D}\hat{\mathcal{Z}})_{ij}+\Big[(\hat{L}\hat{\mathcal{Z}})_{ij}-(\hat{L}\hat{\mathcal{Z}})_{ii}\Big]
\end{equation}
\end{lemma}
{\bf Proof.} Below we use order 2 matrix notations for simplicity. We have
\begin{equation}
\hat{L}=\hat{K}-\hat{N}\hat{\Phi}\;\Longrightarrow \; \hat{N}\hat{\Phi}+\hat{L}=\hat{N}+\hat{D}
\end{equation}
Apply $\hat{\mathcal{Z}}$ to both sides, so
\begin{equation}\label{aux11}
\hat{N}\hat{\Phi}\hat{\mathcal{Z}}+\hat{L}\hat{\mathcal{Z}}=\hat{N}\hat{\mathcal{Z}}+\hat{D}\hat{\mathcal{Z}}\;\Longrightarrow \hat{N}(I-\hat{\Phi})\hat{\mathcal{Z}}=\hat{L}\hat{\mathcal{Z}}-\hat{D}\hat{\mathcal{Z}}
\end{equation}
By Lemma \ref{cflema1}, $(I-\hat{\Phi})\hat{\mathcal{Z}}=I-\hat{\Omega}$, so
$\hat{N}(I-\hat{\Phi})\hat{\mathcal{Z}}=\hat{N}(I-\hat{\Omega})=\hat{N}-\hat{N}\hat{\Omega}$, and by (\ref{aux11}) we get
\begin{equation}
\hat{N}-\hat{N}\hat{\Omega}=\hat{L}\hat{\mathcal{Z}}-\hat{D}\hat{\mathcal{Z}}\;\Longrightarrow\;\hat{N}=\hat{L}\hat{\mathcal{Z}}-\hat{D}\hat{\mathcal{Z}}+\hat{N}\hat{\Omega}
\end{equation}
Now note that the product $\hat{N}\hat{\Omega}$ is of the form
\begin{equation}\label{dim2e1}
\hat{N}\hat{\Omega}=\begin{bmatrix} 0 & \hat{k}_{12} \\ \hat{k}_{21} & 0 \end{bmatrix}\begin{bmatrix} \hat{\Omega}_{11} & \hat{\Omega}_{11} \\ \hat{\Omega}_{11} & \hat{\Omega}_{11} \end{bmatrix}=\begin{bmatrix} \hat{k}_{12}\hat{\Omega}_{11} & \hat{k}_{12}\hat{\Omega}_{11} \\ \hat{k}_{21}\hat{\Omega}_{11} & \hat{k}_{21}\hat{\Omega}_{11}\end{bmatrix},
\end{equation}
so we can write $(\hat{N}\hat{\Omega})_{ij}=(\hat{N}\hat{\Omega})_{i}$, that is, only the row choice matters (it is clear that this fact on expression (\ref{dim2e1}) is also valid for any number of sites). Therefore from
\begin{equation}
\hat{N}=\hat{L}\hat{\mathcal{Z}}-\hat{D}\hat{\mathcal{Z}}+\hat{N}\hat{\Omega}
\end{equation}
we get
$$\hat{N}_{ij}=(\hat{L}\hat{\mathcal{Z}})_{ij}-(\hat{D}\hat{\mathcal{Z}})_{ij}+(\hat{N}\hat{\Omega})_i$$
Then if $i=j$ we get $0=\hat{N}_{ij}=(\hat{L}\hat{\mathcal{Z}})_{jj}-(\hat{D}\hat{\mathcal{Z}})_{jj}+(\hat{N}\hat{\Omega})_j$ from which we get $(\hat{N}\hat{\Omega})_i=(\hat{D}\hat{\mathcal{Z}})_{ii}-(\hat{L}\hat{\mathcal{Z}})_{ii}$. Finally, for $i\neq j$, we get
$$\hat{N}_{ij}=(\hat{L}\hat{\mathcal{Z}})_{ij}-(\hat{D}\hat{\mathcal{Z}})_{ij}+(\hat{N}\hat{\Omega})_i=(\hat{L}\hat{\mathcal{Z}})_{ij}-(\hat{D}\hat{\mathcal{Z}})_{ij}+(\hat{D}\hat{\mathcal{Z}})_{ii}-(\hat{L}\hat{\mathcal{Z}})_{ii}$$
\begin{equation}\label{quantum_aldous}
=(\hat{D}\hat{\mathcal{Z}})_{ii}-(\hat{D}\hat{\mathcal{Z}})_{ij}+\Big[(\hat{L}\hat{\mathcal{Z}})_{ij}-(\hat{L}\hat{\mathcal{Z}})_{ii}\Big],
\end{equation}

\qed

Now we note that $\hat{\mathcal{Z}}$ satisfies
\begin{equation}\label{zpreserves}
Tr(\hat{\mathcal{Z}}\rho)=Tr\Big((\Big[I+\sum_{n\geq 1}(\hat{\Phi}^n-\hat{\Omega})\Big]\rho\Big)=Tr(\rho)+\sum_{n\geq 1}\Big[Tr(\hat{\Phi}^n \rho)-Tr(\hat{\Omega}\rho)\Big]=Tr(\rho)
\end{equation}
as $\Phi$ and $\Omega$ are trace-preserving. Then we perform a calculation concerning the product $\hat{L}\hat{\mathcal{Z}}$. For simplicity we consider the case of 2 sites below. We have
$$\hat{L}{\hat{\mathcal{Z}}}\begin{bmatrix} \rho \\ 0 \end{bmatrix}=\begin{bmatrix} \hat{L}_{11} & \hat{L}_{12} \\ \hat{L}_{21} & \hat{L}_{22} \end{bmatrix}\begin{bmatrix} \hat{Z}_{11} & \hat{Z}_{12} \\ \hat{Z}_{21} & \hat{Z}_{22}\end{bmatrix}\begin{bmatrix} \rho \\ 0 \end{bmatrix}=\begin{bmatrix}\hat{L}_{11} & \hat{L}_{12} \\ \hat{L}_{21} & \hat{L}_{22} \end{bmatrix}\begin{bmatrix} \hat{Z}_{11}\rho \\ \hat{Z}_{21}\rho\end{bmatrix}=\begin{bmatrix} \hat{L}_{11}\hat{Z}_{11}\rho + \hat{L}_{12}\hat{Z}_{21}\rho \\ \hat{L}_{21}\hat{Z}_{11}\rho+\hat{L}_{22}\hat{Z}_{21}\rho \end{bmatrix}$$
\begin{equation}\label{lz_eq1}
=\begin{bmatrix} (\hat{L}_{11}\hat{Z}_{11} + \hat{L}_{12}\hat{Z}_{21})\rho \\ (\hat{L}_{21}\hat{Z}_{11}+\hat{L}_{22}\hat{Z}_{21})\rho \end{bmatrix}=\begin{bmatrix} (\hat{L}\hat{\mathcal{Z}})_{11}\rho \\ (\hat{L}\hat{\mathcal{Z}})_{21}\rho\end{bmatrix}
\end{equation}
Similarly,
$$\hat{L}{\hat{\mathcal{Z}}}\begin{bmatrix} 0 \\ \rho  \end{bmatrix}=\begin{bmatrix} \hat{L}_{11} & \hat{L}_{12} \\ \hat{L}_{21} & \hat{L}_{22} \end{bmatrix}\begin{bmatrix} \hat{Z}_{11} & \hat{Z}_{12} \\ \hat{Z}_{21} & \hat{Z}_{22}\end{bmatrix}\begin{bmatrix} 0 \\\rho \end{bmatrix}=\begin{bmatrix}\hat{L}_{11} & \hat{L}_{12} \\ \hat{L}_{21} & \hat{L}_{22} \end{bmatrix}\begin{bmatrix} \hat{Z}_{12}\rho \\ \hat{Z}_{22}\rho\end{bmatrix}=\begin{bmatrix} \hat{L}_{11}\hat{Z}_{12}\rho + \hat{L}_{12}\hat{Z}_{22}\rho \\ \hat{L}_{21}\hat{Z}_{12}\rho+\hat{L}_{22}\hat{Z}_{22}\rho \end{bmatrix}$$
\begin{equation}\label{lz_eq2}
=\begin{bmatrix} (\hat{L}_{11}\hat{Z}_{12} + \hat{L}_{12}\hat{Z}_{22})\rho \\ (\hat{L}_{21}\hat{Z}_{12}+\hat{L}_{22}\hat{Z}_{22})\rho \end{bmatrix}=\begin{bmatrix} (\hat{L}\hat{\mathcal{Z}})_{12}\rho \\ (\hat{L}\hat{\mathcal{Z}})_{22}\rho\end{bmatrix}
\end{equation}
The matrix calculations for $k>2$ sites are similar.

\medskip

{\bf Proof of Theorem \ref{quantum_aldous3}}. We begin with
\begin{equation}
\hat{N}_{ij}=(\hat{D}\hat{\mathcal{Z}})_{ii}-(\hat{D}\hat{\mathcal{Z}})_{ij}+\Big[(\hat{L}\hat{\mathcal{Z}})_{ij}-(\hat{L}\hat{\mathcal{Z}})_{ii}\Big]
\end{equation}
By Lemma \ref{very_usedlater}, if $\rho$ is a density matrix concentrated on site $j$ then for all $i$ and all $c\in\mathbb{R}$, we have $Tr(\hat{L}_{ij}c\rho)=Tr(c\rho)=c$. Also note that $\hat{Z}_{ij}\rho$ is positive semidefinite (thus implying that it is a multiple of a density matrix). In particular, by using (\ref{lz_eq1}), (\ref{lz_eq2}) and the fact that $\hat{\mathcal{Z}}$ is trace preserving, we have for $\rho$ positive semidefinite that
$$Tr((\hat{L}\hat{\mathcal{Z}})_{11}\rho)=Tr((\hat{L}_{11}\hat{Z}_{11} + \hat{L}_{12}\hat{Z}_{21})\rho)=Tr(\hat{L}_{11}\hat{Z}_{11}\rho)+Tr(\hat{L}_{12}\hat{Z}_{21}\rho)$$
\begin{equation}
=Tr(\hat{Z}_{11}\rho)+Tr(\hat{Z}_{21}\rho)=Tr\Big((\hat{\mathcal{Z}}\begin{bmatrix} \rho \\ 0 \end{bmatrix}\Big)=Tr(\rho)
\end{equation}
and similarly
$$Tr((\hat{L}\hat{\mathcal{Z}})_{12}\rho)=Tr((\hat{L}_{11}\hat{Z}_{12} + \hat{L}_{12}\hat{Z}_{22})\rho)=Tr(\hat{L}_{11}\hat{Z}_{12}\rho)+Tr(\hat{L}_{12}\hat{Z}_{22}\rho)$$
\begin{equation}
=Tr(\hat{Z}_{12}\rho)+Tr(\hat{Z}_{22}\rho)=Tr\Big((\hat{\mathcal{Z}}\begin{bmatrix} 0 \\ \rho \end{bmatrix}\Big)=Tr(\rho)
\end{equation}
We conclude that $Tr((\hat{L}\hat{\mathcal{Z}})_{11}\rho)=Tr((\hat{L}\hat{\mathcal{Z}})_{12}\rho)$ and the result follows. The proof that $Tr((\hat{L}\hat{\mathcal{Z}})_{21}\rho)=Tr((\hat{L}\hat{\mathcal{Z}})_{22}\rho)$ is identical. The proof is independent of $n$ and the general case of $k>2$ follows from the corresponding versions of eqs. (\ref{lz_eq1}) and (\ref{lz_eq2}).

\qed

\begin{example}\label{numexample}
Consider the following OQW with degree of freedom $n=2$ acting on $k=2$ sites,
\begin{equation}
\hat{\Phi}=\begin{bmatrix} B_{11} & B_{12} \\ B_{21} & B_{22}\end{bmatrix},\; B_{11}=\frac{1}{2}I,\; B_{12}=\frac{\sqrt{3}}{2}I,\; B_{21}=\frac{\sqrt{3}}{2}\begin{bmatrix} 0 & 1 \\ 1 & 0 \end{bmatrix},\; B_{22}=\frac{i}{2}\begin{bmatrix} 0 & -1 \\ 1 & 0 \end{bmatrix}
\end{equation}
Then
\begin{equation}
[\Phi]=\frac{1}{4}\begin{bmatrix} 1 & 0 & 0 & 0 & 3 & 0 & 0 & 0 \\ 0 & 1 & 0 & 0 & 0 & 3 & 0 & 0 \\0 & 0 & 1 & 0 & 0 & 0 & 3 & 0 \\ 0 & 0 & 0 & 1 & 0 & 0 & 0 & 3 \\ 0 & 0 & 0 & 3 & 0 & 0 & 0  & 1 \\ 0 & 0 & 3 & 0 & 0 & 0 & -1 &  0 \\0 & 3 & 0 & 0 & 0 & -1 & 0 & 0 \\3 & 0 & 0 & 0 & 1 & 0 & 0 & 0 \end{bmatrix}
\end{equation}
It is a simple matter to show that $\Phi$ has 1 as the unique eigenvalue on the unit circle, its associated eigenspace is 1-dimensional, generated by $[I \; I]^T$ and that its asymptotic limit is the map $\hat{\Omega}$. We would like to check that for $\rho=(\rho)_{ij}\in M_2(\mathbb{C})$ density matrix we have
\begin{equation}\label{check1}
Tr(\hat{N}_{12}\rho)=Tr((\hat{D}\hat{\mathcal{Z}})_{11}\rho)-Tr((\hat{D}\hat{\mathcal{Z}})_{12}\rho)
\end{equation}
\begin{equation}\label{check2}
Tr(\hat{N}_{21}\rho)=Tr((\hat{D}\hat{\mathcal{Z}})_{22}\rho)-Tr((\hat{D}\hat{\mathcal{Z}})_{21}\rho)
\end{equation}
First note that for any $B_{ij}$ order 2 matrices acting on 2 sites, the mean hitting operator $\hat{K}$ is quite simple:
\begin{equation}\label{eq_k11}
\hat{k}_{11}=M_{B_{11}}+\sum_{r=2}^\infty rM_{B_{12}}M_{B_{22}}^{r-2}M_{B_{21}}=M_{B_{11}}+M_{B_{12}}\Big(\sum_{r=2}^\infty rM_{B_{22}}^{n-2}\Big)M_{B_{21}}
\end{equation}
For instance, the term $M_{B_{12}}M_{B_{22}}^{r-2}M_{B_{21}}$ above means that (read indices from right to left) if one begins at 1 and first returns to 1 at time $r$, then it must spend $r-2$ units of time at site 2.
Similarly,
\begin{equation}\label{eq_k22}
\hat{k}_{22}=M_{B_{22}}+M_{B_{21}}\Big(\sum_{r=2}^\infty nM_{B_{11}}^{n-2}\Big)M_{B_{12}}
\end{equation}
\begin{equation}\label{eq_k12_21}
\hat{k}_{12}=\sum_{r=1}^\infty rM_{B_{12}}M_{B_{22}}^{r-1}=M_{B_{12}}\sum_{r=1}^\infty rM_{B_{22}}^{r-1},\;\;\;\;\;\;\hat{k}_{21}=M_{B_{21}}\sum_{r=1}^\infty rM_{B_{11}}^{r-1}
\end{equation}
For the example, a calculation gives (some decimals have been omitted for simplicity):
$$\hat{k}_{11}=\begin{bmatrix} 0.72 & 0 & 0 & 1.28 \\ 0 & -0.22 & 1.28 & 0 \\ 0 & 1.28 & -0.22 & 0 \\ 1.28 & 0 & 0 & 0.72\end{bmatrix},\;\;\;\hat{k}_{12}=\begin{bmatrix} 0.90667 & 0 & 0 & 0.426667 \\ 0 & 0.906667 & -0.426667 & 0 \\ 0 & -0.42667 & 0.906667 & 0 \\ 0.426667 & 0 & 0 & 0.906667\end{bmatrix},\;\;\;$$
\begin{equation}
\hat{k}_{21}=\begin{bmatrix} 0 & 0 & 0 & 1.333 \\ 0 & 0 & 1.333 & 0 \\ 0 & 1.333 & 0 & 0 \\ 1.333 & 0 & 0 & 0 \end{bmatrix},\;\;\;\hat{k}_{22}=\begin{bmatrix} 0 & 0 & 0 & 2 \\ 0 & 0 & 1.5 & 0 \\ 0 & 1.5 & 0 & 0 \\ 2 & 0 & 0 & 0 \end{bmatrix}
\end{equation}
\begin{equation}
\mathcal{Z}=\begin{bmatrix} 0.8333 & 0 & 0 &0 & 0.333 & 0 & 0 & -0.16667 \\ 0 & 2 & 1.333 & 0 & 0 & 1.333 & 0.6667 & 0 \\ 0 & 1.333 & 2 & 0 & 0 & 0.6667 & 1.333 & 0 \\ 0 & 0 & 0 & 0.8333 & -0.16667 & 0 & 0 & 0.333 \\ -0.16667 & 0 & 0 & 0.333 & 0.6667 & 0 & 0 & 0.16667 \\ 0 & 0.6667 & 1.333 & 0 & 0 & 1.333 & 0.6667 & 0 \\ 0 & 1.333 & 0.6667 & 0 & 0 & 0.6667 & 1.333 & 0 \\ 0.333 & 0 & 0 & -0.16667 & 0.16667 & 0 & 0 & 0.6667 \end{bmatrix}
\end{equation}
In particular, $Tr(\hat{k}_{11}\rho)=Tr(\hat{k}_{22}\rho)=2$. Also,
\begin{equation}
Tr(\hat{N}_{12}\rho)=Tr(\hat{k}_{12}\rho))=1.3333(\rho_{11}+\rho_{22})=1.3333
\end{equation}
As for the right side, we calculate
\begin{equation}
\hat{D}\hat{\mathcal{Z}}=\begin{bmatrix} 0.5999 & 0 & 0 & 1.0666 & 0.02666 & 0 & 0 & 0.30666 \\ 0 & 1.26666 & 2.26666 & 0 & 0 & 0.5600001 & 1.5599 & 0 \\ 0 & 2.2666 & 1.2666 & 0 & 0 & 1.559999 & 0.5600001  & 0 \\ 1.0666 & 0 & 0 & 0.5999 & 0.30666 & 0 & 0 & 0.02666 \\ 0.6666 & 0 & 0 & -0.3333 & 0.3333 & 0 & 0 & 1.3333 \\ 0 & 2 &  1 & 0 & 0 & 1 & 2 & 0 \\ 0 & 1 & 2 & 0 & 0 & 2 & 1 & 0 \\ -0.3333 & 0 & 0 & 0.66666 &  1.3333 & 0 & 0 & 0.3333 \end{bmatrix}
\end{equation}
Then
\begin{equation}
Tr(\hat{D}\hat{\mathcal{Z}}_{11}\rho)=1.6666(\rho_{11}+\rho_{22})=1.6666
\end{equation}
and
\begin{equation}
Tr(\hat{D}\hat{\mathcal{Z}}_{12}\rho)=0.3333(\rho_{11}+\rho_{22})=0.3333
\end{equation}
which is consistent with (\ref{check1}). We proceed similarly with (\ref{check2}) obtaining the results, $Tr(\hat{N}_{21}\rho)=1.333$, $Tr(\hat{D}\hat{\mathcal{Z}}_{22}\rho)=1.6666$, $Tr(\hat{D}\hat{\mathcal{Z}}_{21}\rho)=0.3333$.   
\end{example}

\qee

\section{Proof of Corollary \ref{rtlemma}}\label{sec77}

The idea of the proof is the same as for the classical case. We recall that in the classical theory of Markov chains, if $Z$ is the fundamental matrix given in the introduction, then
\begin{equation}
\sum_j Z_{ij}=0,\;\;\;\forall \;i
\end{equation}
From the mean hitting time formula $\pi_jE_iT_j=Z_{jj}-Z_{ij}$, if we sum over $j$ we get
\begin{equation}
\sum_j\pi_jE_iT_j=\sum_j Z_{jj}
\end{equation}
that is, the left hand side does not depend on $i$. Now recall the definition of $\mathcal{Z}$ for ergodic OQWs and write the Mean Hitting Time Formula for OQWs:
\begin{equation}\label{final_calc}
Tr((\hat{D}^{-1}\hat{N})_{ij}\rho)=Tr(\hat{\mathcal{Z}}_{ii}\rho)-Tr(\hat{\mathcal{Z}}_{ij}\rho)
\end{equation}
The trace preservation of $\hat{\mathcal{Z}}$, eq.(\ref{zpreserves}), implies that
\begin{equation}
\sum_i Tr(\hat{Z}_{ij}\rho)=1
\end{equation}
and so by summing in $i$ both sides of (\ref{final_calc}) we get that
\begin{equation}\label{rtlemma0}
t_{\odot}(\rho):=\sum_i Tr((\hat{D}^{-1}\hat{N})_{ij}\rho)=[\sum_i Tr(\hat{\mathcal{Z}}_{ii}\rho)]-1
\end{equation}
does not depend on $j$.

\begin{example}\label{carboneex}
Let
\begin{equation}
L=\frac{1}{\sqrt{3}}\begin{bmatrix} 1 & 1 \\ 0 & 1 \end{bmatrix},\;\;\;R=\frac{1}{\sqrt{3}}\begin{bmatrix} 1 & 0 \\ -1 & 1 \end{bmatrix}
\end{equation}
By [\cite{carbone}, Theorem 9.6], the OQW given by
\begin{equation}
[\Phi]=\begin{bmatrix} [0] & [R] & [L] \\ [L] & [0] & [R] \\ [R] & [L] & [0]\end{bmatrix}
\end{equation}
is aperiodic and irreducible and so $\Phi\in\mathcal{E}$. Even for 3 sites the calculations are already quite long (the matrix representation for $\mathcal{Z}$ in this case has order 12), but a computer algorithm can easily perform the computation of $\mathcal{Z}$ and the subsequence target times. One such calculation produces, for any order 2 density matrix $\rho=(\rho_{ij})$,
\begin{equation}
Tr(\hat{\mathcal{Z}}_{ii}\rho)=.717948\rho_{11}+.717948\rho_{22}=.717948,\;\;\;i=1,2,3
\end{equation}
and so
\begin{equation}
t_{\odot}(\rho)=[\sum_i Tr(\hat{\mathcal{Z}}_{ii}\rho)]-1=1.153844
\end{equation}

\end{example}

\qee

\section{Proof of Theorem \ref{chzhthm}}\label{sec6}

In this section we will prove a mean hitting time formula for finite ergodic OQWs, inspired by the result due to H. Chen and F. Zhang \cite{chenz} for irreducible Markov chains. Such proof consists of 3 preliminary results, followed by the main theorem. We will follow the same strategy, pointing out similarities and providing the corresponding proof for matrix representations of CPT maps when required.

\medskip

By the {\bf minimal polynomial} of an ergodic OQW $\Phi$ we mean the one associated to its matrix representation. That is, the monic polynomial $p$ of least degree such that $p([\Phi])=0$. Since $1$ is an eigenvalue of multiplicity 1, we see that $p$ is of the form $p(x)=(x-1)f(x)$ and we write
\begin{equation}
f(x)=x^r+a_1x^{r-1}+a_2x^{r-2}+\cdots+a_{r-1}x+a_r
\end{equation}
In \cite{chenz} it has been shown that the row vectors of the matrix $f(P)$ are similar. The result is also true in the case of matrix representation of ergodic OQWs (in the place of irreducible stochastic matrices), the reasoning being almost the same as the one presented in \cite{chenz}. For completeness we describe here the simple adaptation of the proof.

\begin{lemma}\label{chenzlema1} The row vectors of $f([\Phi])$ are similar. \end{lemma}
{\bf Proof.} As $p(x)=(x-1)f(x)$ is the minimal polynomial of $[\Phi]$ we obtain $[\Phi]f([\Phi])-f([\Phi])=0$, that is, $[\Phi]f([\Phi])=f([\Phi])$, so each column of $f([\Phi])$ is an eigenvector of $\Phi$ corresponding to the eigenvalue 1. Being ergodic, each column of $f([\Phi])$ must be equal to a multiple of
\begin{equation}
u=vec([I\;I\;\cdots \;I]^T),\;\;\;I=I_n
\end{equation}
The lemma follows.
\qed

\begin{remark} If in the proof of the above Lemma we have, for instance, $k=3$ sites and $n=2$ then
\begin{equation}
u=vec([I_2\;I_2\;I_2]^T)=
[1 \; 0 \; 0 \; 1 \; 1 \; 0 \; 0 \; 1 \; 1 \; 0 \; 0 \; 1]^T
\end{equation}
\end{remark}

Now we state the second lemma, noting that the result holds for any polynomial.

\begin{lemma}\label{chenzlema2}\cite{chenz} Let $f(x)=x^r+a_1x^{r-1}+a_2x^{r-2}+\cdots+a_{r-1}x+a_r$, and
\begin{equation}
\alpha_0=(-a_1,-a_2,\dots,-a_r),\;\;\; \beta=(x^{r-1},x^{r-2},\dots,x,1)^T
\end{equation}
Then for any $m\geq 0$ there exists a polynomial $q_m(x)$ of degree $m$ and a row vector $\alpha_m=(\alpha_{m,1},\alpha_{m,2},\dots,\alpha_{m,r})$ such that
\begin{equation}
x^{r+m}=q_m(x)f(x)+\alpha_m\beta,\;\;\;\alpha_m=\alpha_0M^m,
\end{equation}
and
\begin{equation}
M=\begin{bmatrix} -a_1 & -a_2 & \cdots & -a_{r-1} & -a_r \\ 1 & 0 & \cdots & 0 & 0 \\ 0 & 1 & \cdots & 0 & 0 \\ \vdots & \vdots & \ddots & \vdots & \vdots \\ 0 & 0 & \cdots & 1 & 0 \end{bmatrix}
\end{equation}
\end{lemma}

Now we state the third lemma in terms of matrix representations of quantum channels.

\begin{lemma}\label{lemma3rd}
Let $\Phi$ be a finite ergodic OQW, let $(x-1)f(x)$ denote the minimal polynomial of $\hat{\Phi}$ and $f(x)=a_0x^r+a_1x^{r-1}+a_2x^{r-2}+\cdots+a_{r-1}x+a_r$. Then
\begin{equation}
\sum_{m=0}^\infty \hat{\Phi}^{r+m}=\sum_{n=0}^{r-1}b_n\hat{\Phi}^n+B,\;\;\;b_n=-\frac{\sum_{l=r-n}^r a_l}{f(1)}
\end{equation}
where $B$ is a matrix with the same row vectors.
\end{lemma}
{\bf Proof.} By replacing the stochastic matrix with the matrix representation of the ergodic OQW, the proof is a simple adaptation of [\cite{chenz}, Lemma 2.4] together with Lemmas \ref{chenzlema1} and \ref{chenzlema2}.
\qed

{\bf Proof of Theorem \ref{chzhthm}.} We begin with expression (\ref{aldous_eqq1}), the Mean Hitting Time Formula for OQWs, so we can write
$$Tr(\hat{N}_{ij}\rho)=Tr((\hat{D}\hat{\mathcal{Z}})_{ii}\rho)-Tr((\hat{D}\hat{\mathcal{Z}})_{ij}\rho)$$
$$=Tr\Big[\Big(\hat{D}\hat{I}+\sum_{s=1}^\infty \hat{D}(\hat{\Phi}^s-\hat{\Omega})\Big)_{ii}\rho\Big]-Tr\Big[\Big(\hat{D}\hat{I}+\sum_{s=1}^\infty \hat{D}(\hat{\Phi}^s-\hat{\Omega})\Big)_{ij}\rho\Big]$$
\begin{equation}
=Tr\Big[\hat{D}_{ii}\rho-\hat{D}_{ij}\rho+\sum_{s=1}^\infty\Big((\hat{D}\hat{\Phi}^s)_{ii}\rho-(\hat{D}\hat{\Phi}^s)_{ij}\rho
-(\hat{D}\hat{\Omega})_{ii}\rho+(\hat{D}\hat{\Omega})_{ij}\rho\Big)\Big]
\end{equation}
Since all blocks of $\Omega$ are equal, the terms with $\Omega$ vanish and by changing the index sum we obtain
\begin{equation}
Tr(\hat{N}_{ij}\rho)=Tr\Big[\sum_{s=0}^\infty\Big((\hat{D}\hat{\Phi}^s)_{ii}\rho-(\hat{D}\hat{\Phi}^s)_{ij}\rho
\Big)\Big]=\sum_{s=0}^\infty Tr\Big[\Big((\hat{D}\hat{\Phi}^s)_{ii}-(\hat{D}\hat{\Phi}^s)_{ij}\Big)\rho
\Big]
\end{equation}
Note that $\hat{D}=diag(\hat{k}_{11},\dots,\hat{k}_{22})$, so we can also write
\begin{equation}
Tr(\hat{N}_{ij}\rho)=Tr\Big[\sum_{s=0}^\infty (\hat{k}_{ii}(\hat{\Phi}^s)_{ii}-\hat{k}_{ii}(\hat{\Phi}^s)_{ij})\rho\Big],
\end{equation}
where above we used that fact that if $D=diag(d_{11},\dots,d_{kk})$ and $A$ is any matrix then $(DA)_{ij}=d_{ii}A_{ij}$. Now if $\hat{A}$ is a block matrix with order $k$ blocks $\hat{A}_{ij}$, let $e_i=[0 \cdots  0 \; I_k \; 0\cdots 0]$ where $I_k$ appears in the $i$-th position, $i=1,\dots,n$. Then we clearly have
\begin{equation}
\hat{A}_{ij}=e_i\hat{A}e_j^T,\;\;\;i,j=1,\dots,n
\end{equation}
With this in mind we write, using Lemma \ref{lemma3rd},
\begin{equation}
\sum_{s=0}^\infty\Big((\hat{\Phi}^s)_{ii}-(\hat{\Phi}^s)_{ij}\Big)=e_i\Big(\sum_{s=0}^\infty\hat{\Phi}^s\Big)e_i^T-e_i\Big(\sum_{s=0}^\infty\hat{\Phi}^s\Big)e_j^T
\end{equation}
$$=e_i\Big(\sum_{s=0}^{r-1}\hat{\Phi}^s\Big)e_i^T-e_i\Big(\sum_{s=0}^{r-1}\hat{\Phi}^s\Big)e_j^T+
e_i\Big(\sum_{m=0}^\infty\hat{\Phi}^{r+m}\Big)e_i^T-e_i\Big(\sum_{m=0}^\infty\hat{\Phi}^{r+m}\Big)e_j^T$$
\begin{equation}
=e_i\Big(\sum_{s=0}^{r-1}\hat{\Phi}^s\Big)e_i^T-e_i\Big(\sum_{s=0}^{r-1}\hat{\Phi}^s\Big)e_j^T
+e_iBe_i^T-e_iBe_j^T+e_i\Big(\sum_{s=0}^{r-1}b_s\hat{\Phi}^s\Big)e_i^T-e_i\Big(\sum_{s=0}^{r-1}b_s\hat{\Phi}^s\Big)e_j^T
\end{equation}
$$=e_i\Big(\sum_{s=0}^{r-1}(1+b_s)\hat{\Phi}^s\Big)e_i^T-e_i\Big(\sum_{s=0}^{r-1}(1+b_s)\hat{\Phi}^s \Big)e_j^T=e_i\Big(\sum_{s=0}^{r-1}\frac{\sum_{l=0}^{r-s-1}a_l}{f(1)}\hat{\Phi}^s\Big)e_i^T-e_i\Big(\sum_{s=0}^{r-1}\frac{\sum_{l=0}^{r-s-1}a_l}{f(1)}\hat{\Phi}^s\Big)e_j^T$$
\begin{equation}
=\frac{1}{f(1)}\sum_{s=0}^{r-1}\sum_{l=0}^{r-s-1}a_l((\hat{\Phi}^s)_{ii}-(\hat{\Phi}^s)_{ij})
\end{equation}
\qed

\begin{example}
Consider once again Example \ref{numexample}. The minimal polynomial for the associated block matrix is
\begin{equation}
p(x)=x^8-x^7+\frac{1}{4}x^6+\frac{1}{16}x^5-\frac{43}{64}x^4+\frac{41}{128}x^3-\frac{5}{64}x^2+\frac{5}{256}x+\frac{25}{256}
\end{equation}
If $p(x)=(x-1)f(x)$ then
\begin{equation}
f(x)=x^7+\frac{1}{4}x^5+\frac{5}{16}x^4-\frac{23}{64}x^3-\frac{5}{128}x^2-\frac{15}{128}x-\frac{25}{256}
\end{equation}
which gives $f(1)=243/256$. With the previous calculations we obtain the same as before, namely $Tr(\hat{N}_{12}\rho)=1.3333$.
\end{example}
\qee

\bigskip

{\bf Acknowledgements.}  The author is grateful to an anonymous referee for many useful comments and suggestions that led to a marked improvement of the paper, and to F.A. Gr\"unbaum for several discussions on hitting times in an open quantum setting.

\end{document}